%% file: paper.tex
\documentclass[acmsmall]{acmart}

\acmJournal{PACMPL}
\acmVolume{1}
\acmNumber{CONF} % CONF = POPL or ICFP or OOPSLA
\acmArticle{1}
\acmYear{2018}
\acmMonth{1}
\acmDOI{} % \acmDOI{10.1145/nnnnnnn.nnnnnnn}
\startPage{1}

\setcopyright{none}
\usepackage{booktabs}   %% For formal tables:
\usepackage{subcaption} %% For complex figures with subfigures/subcaptions
\usepackage{adjustbox}
\usepackage{arydshln}
\usepackage{circledsteps}
\usepackage{enumitem}
\usepackage{fancybox}
\usepackage{makecell}
\usepackage{multirow}
\usepackage{listings}
\usepackage{spreadtab}
\usepackage{tikz}
\usepackage{ulem}
\usepackage{todonotes}
\usepackage{xcolor}
\usepackage{xspace}
\usetikzlibrary{calc, fit, tikzmark, patterns, shapes.multipart, positioning, arrows.meta, matrix,decorations.pathmorphing}
\pgfdeclarepatternformonly{wide north east lines}{\pgfqpoint{-1pt}{-1pt}}{\pgfqpoint{9pt}{9pt}}{\pgfqpoint{8pt}{8pt}}%
{
  \pgfsetlinewidth{1pt}
  \pgfpathmoveto{\pgfqpoint{0pt}{0pt}}
  \pgfpathlineto{\pgfqpoint{8pt}{8pt}}
  \pgfpathmoveto{\pgfqpoint{7pt}{-1pt}}
  \pgfpathlineto{\pgfqpoint{9pt}{1pt}}
  \pgfpathmoveto{\pgfqpoint{-1pt}{7pt}}
  \pgfpathlineto{\pgfqpoint{1pt}{9pt}}
  \pgfusepath{stroke}
}

\pgfdeclarepatternformonly{wide north west lines}{\pgfqpoint{-1pt}{-1pt}}{\pgfqpoint{9pt}{9pt}}{\pgfqpoint{8pt}{8pt}}%
{
  \pgfsetlinewidth{1pt}
  \pgfpathmoveto{\pgfqpoint{0pt}{8pt}}
  \pgfpathlineto{\pgfqpoint{8pt}{0pt}}
  \pgfpathmoveto{\pgfqpoint{7pt}{9pt}}
  \pgfpathlineto{\pgfqpoint{9pt}{7pt}}
  \pgfpathmoveto{\pgfqpoint{-1pt}{1pt}}
  \pgfpathlineto{\pgfqpoint{1pt}{-1pt}}
  \pgfusepath{stroke}
}

\input{macros}
\input{statmacros}
\begin{document}

%% Title information
\title{Resource Specifications for Resource-Manipulating Programs}         %% [Short Title] is optional;
                                        %% when present, will be used in
                                        %% header instead of Full Title.
%\titlenote{with title note}             %% \titlenote is optional;
                                        %% can be repeated if necessary;
                                        %% contents suppressed with 'anonymous'
%\subtitle{Subtitle}                     %% \subtitle is optional
%\subtitlenote{with subtitle note}       %% \subtitlenote is optional;
                                        %% can be repeated if necessary;
                                        %% contents suppressed with 'anonymous'

%% Author information
%% Contents and number of authors suppressed with 'anonymous'.
%% Each author should be introduced by \author, followed by
%% \authornote (optional), \orcid (optional), \affiliation, and
%% \email.
%% An author may have multiple affiliations and/or emails; repeat the
%% appropriate command.
%% Many elements are not rendered, but should be provided for metadata
%% extraction tools.

%% Author with single affiliation.
\author{Zachary Grannan}
%\authornote{with author1 note}          %
% \orcid{nnnn-nnnn-nnnn-nnnn}             %
\affiliation{
  %\position{Position1}
  %\department{Department1}              %
  \institution{University of British Columbia}            %
  % \streetaddress{Street1 Address1}
  \city{Vancouver}
  % \state{State1}
  % \postcode{Post-Code1}
  \country{Canada}                    %
}
% \email{first1.last1@inst1.edu}          %

\author{Alexander J. Summers}
% \authornote{with author2 note}          %
% \orcid{nnnn-nnnn-nnnn-nnnn}             %
\affiliation{
%   \position{Position2a}
%   \department{Department2a}             %
   \institution{University of British Columbia}           %
%   \streetaddress{Street2a Address2a}
   \city{Vancouver}
%   \state{State2a}
%   \postcode{Post-Code2a}
   \country{Canada}                   %
}

%% Abstract
%% Note: \begin{abstract}...\end{abstract} environment must come
%% before \maketitle command
\begin{abstract}
\input{abstract}
\end{abstract}

%% 2012 ACM Computing Classification System (CSS) concepts
%% Generate at 'http://dl.acm.org/ccs/ccs.cfm'.
\begin{CCSXML}
<ccs2012>
<concept>
<concept_id>10011007.10011006.10011008</concept_id>
<concept_desc>Software and its engineering~General programming languages</concept_desc>
<concept_significance>500</concept_significance>
</concept>
<concept>
<concept_id>10003456.10003457.10003521.10003525</concept_id>
<concept_desc>Social and professional topics~History of programming languages</concept_desc>
<concept_significance>300</concept_significance>
</concept>
</ccs2012>
\end{CCSXML}

\ccsdesc[500]{Software and its engineering~General programming languages}
\ccsdesc[300]{Social and professional topics~History of programming languages}
%% End of generated code

%% Keywords
%% comma separated list
%\keywords{keyword1, keyword2, keyword3}  %% \keywords are mandatory in final camera-ready submission

%% \maketitle
%% Note: \maketitle command must come after title commands, author
%% commands, abstract environment, Computing Classification System
%% environment and commands, and keywords command.
\maketitle

\input{introduction}
\input{motivation}
\input{methodology}
\input{implementation}

\input{casestudy}
\input{evaluation}
\input{related}
\input{conclusion}

%% Bibliography
\bibliography{bib}

%% Appendix
% \appendix
% \section{Appendix}

% Text of appendix \ldots

\end{document}

%% file: macros.tex
\newcommand\noeditcolour{}
\newcommand{\RSE}{Resource State Expression}

\newcommand{\rse}{resource state expression}
\newcommand{\ink}{\texttt{ink!}}
\newcommand{\prusti}{Prusti}

\newcommand{\secref}[1]{Sec.~\ref{sec:#1}}
\newcommand{\figref}[1]{Fig.~\ref{fig:#1}}

\newcommand{\transferskwd}{\texttt{resource}}
\newcommand{\transfers}[1]{\texttt{\transferskwd(#1)}}
\newcommand{\resourceann}{\texttt{\#[resource\_kind]}}

\newcommand{\rkind}{resource kind}

\newcommand{\rtypeshort}{type}
\newcommand{\rtype}{resource type}
\newcommand{\rtypes}{\rtype{}s}
\newcommand{\RType}{Resource Type}

\newcommand{\rtypeconshort}{constructor}
\newcommand{\rtypecon}{resource constructor}
\newcommand{\Rtypecon}{Resource constructor}

\newcommand{\predinstfull}{abstract predicate instance}
\newcommand{\predinst}{predicate instantiation}
\newcommand{\exhalekeyword}{\texttt{consume!}}
\newcommand{\inhalekeyword}{\texttt{produce!}}
\newcommand{\holdskeyword}{\texttt{holds}}

\newcommand{\holds}[1]{\texttt{holds(#1)}}

\newcommand{\issue}{issue}

\newcommand{\Issue}{Issue}
\newcommand{\Issues}{Issues}
\newcommand{\erc}[1]{ERC-#1}
\newcommand{\psp}[1]{PSP-#1}
\newcommand{\callout}[2]{
\begin{center}
  \shadowbox{
    \begin{minipage}{0.9\textwidth}
      \textbf{#1}\ #2
    \end{minipage}
  }
\end{center}}

\lstnewenvironment{solidity}[1][1]{%
    \lstset{
    	language=Solidity,
    	style=boxed,
    	basicstyle=\small\ttfamily,
    	escapechar=~,
		firstnumber=#1,
    	tabsize=2,
    }
}{}

\lstdefinelanguage{Rust}
{
	sensitive,
	morekeywords=[1]{
		unsafe,async,await,move,
		use,pub,crate,super,self,Self,mod,
		struct,enum,fn,const,static,let,mut,ref,type,impl,dyn,trait,where,as,
		break,continue,if,else,while,for,loop,match,return,yield,in
	},
	morekeywords=[2]{
		assert!, assert_eq!, debug_assert!, debug_assert_eq!, debug_assert_ne!,
		panic!, unimplemented!, unreachable!,
		print!, println!, vec!, matches!
	},
	morekeywords=[3]{
		pure, trusted, ensures, requires, result, invariant, extern_spec,
		body_invariant!, predicate!,
		forall, old, exists,
		after_expiry, before_expiry, on_expiry, assert_on_expiry
	},
	morekeywords=[4]{
		bool,u8,u16,u32,u64,u128,i8,i16,i32,i64,i128,char,str,usize
	},
	morekeywords=[5]{
		Box, Option, Some, None, Ord, Equal, Less, Greater, Result, Ok, Err,
		T, Tree, Node, Empty, Timestamp, TryFromIntError, Utc, Height,
		MAX
	},
	morekeywords=[5]{
		true, false
	},
	morecomment=[l]{//},
	morecomment=[s]{/*}{*/},
	morecomment=[l]{///},
	morecomment=[s]{/*!}{*/},
	morecomment=[l]{//!},
	string=[b]{"},
	alsoletter={!},
}

\newcommand{\listingsize}{\small} % default size used in various environments
\definecolor{DarkGreen}{RGB}{0,90,0}
\definecolor{DarkBlue}{RGB}{30,30,150}
\definecolor{DarkRed}{RGB}{160,20,10}
\lstdefinestyle{colouredRust}
{
	basicstyle=,
	identifierstyle=,
	keywordstyle=[1]\bfseries\color{DarkBlue},  % [1] reserved keywords
	keywordstyle=[2]\color{DarkGreen}, % [2] macro calls
	keywordstyle=[3]\color{DarkRed},   % [3] Prusti stuff
	keywordstyle=[4]\bfseries\color{DarkBlue},  % [4] primitive types
	keywordstyle=[5]\color{DarkGreen}, % [5] traits, types, variants etc
	keywordstyle=[6]\color{DarkGreen}, % [6] primitive values
	commentstyle=\color[gray]{0.4},
	stringstyle=\color{DarkRed},
	literate={!matches!}{!{\color{DarkGreen}matches!}}1,
	columns=spaceflexible,
	keepspaces=true,
	showspaces=false,
	showtabs=false,
	showstringspaces=true
}

\lstdefinestyle{boxed}{
	style=colouredRust,
	numbers=left,
	firstnumber=auto,
	numberblanklines=true,
	frame=trbL,
	numberstyle=\tiny,
	frame=leftline,
	numbersep=7pt,
	framesep=5pt,
	framerule=10pt,
	xleftmargin=15pt,
	backgroundcolor=\color[gray]{0.97},
	rulecolor=\color[gray]{0.90}
}

\lstnewenvironment{rust}[1][1]{%
    \lstset{
    	language=Rust,
    	style=boxed,
    	basicstyle=\listingsize\ttfamily,
    	escapechar=~,
		firstnumber=#1,
    	tabsize=2,
    }
}{}

\lstnewenvironment{rustmini}[1][1]{%
    \lstset{
    	language=Rust,
    	style=boxed,
    	basicstyle=\scriptsize\ttfamily,
    	escapechar=~,
		firstnumber=#1,
    	tabsize=2,
    }
}{}

\lstnewenvironment{rustgood}[1][1]{%
    \lstset{
    	language=Rust,
    	style=boxed,
    	basicstyle=\listingsize\ttfamily,
        backgroundcolor=\color{lime!25},
    	escapechar=~,
		firstnumber=#1,
    	tabsize=2,
    }
}{}
\lstnewenvironment{rustbad}[1][1]{%
    \lstset{
    	language=Rust,
    	style=boxed,
    	basicstyle=\listingsize\ttfamily,
        backgroundcolor=\color{red!25},
    	escapechar=~,
		firstnumber=#1,
    	tabsize=2,
    }
}{}

\lstnewenvironment{rustnolines}{%
    \lstset{
    	language=Rust,
    	style=boxed,
    	basicstyle=\listingsize\ttfamily,
    	escapechar=~,
		numbers=none,
    	tabsize=2,
    }
}{}

\definecolor{verylightgray}{rgb}{.97,.97,.97}

\lstdefinelanguage{Solidity}{
	keywords=[1]{anonymous, assembly, assert, balance, break, call, callcode, case, catch, class, constant, continue, constructor, contract, debugger, default, delegatecall, delete, do, else, emit, event, experimental, export, external, false, finally, for, function, gas, if, implements, import, in, indexed, instanceof, interface, internal, is, length, library, log0, log1, log2, log3, log4, memory, modifier, new, payable, pragma, private, protected, public, pure, push, require, return, returns, revert, selfdestruct, send, solidity, storage, struct, suicide, super, switch, then, this, throw, true, try, typeof, using, value, view, while, with, addmod, ecrecover, keccak256, mulmod, ripemd160, sha256, sha3}, % generic keywords including crypto operations
	keywordstyle=[1]\color{blue}\bfseries,
	keywords=[2]{address, bool, byte, bytes, bytes1, bytes2, bytes3, bytes4, bytes5, bytes6, bytes7, bytes8, bytes9, bytes10, bytes11, bytes12, bytes13, bytes14, bytes15, bytes16, bytes17, bytes18, bytes19, bytes20, bytes21, bytes22, bytes23, bytes24, bytes25, bytes26, bytes27, bytes28, bytes29, bytes30, bytes31, bytes32, enum, int, int8, int16, int24, int32, int40, int48, int56, int64, int72, int80, int88, int96, int104, int112, int120, int128, int136, int144, int152, int160, int168, int176, int184, int192, int200, int208, int216, int224, int232, int240, int248, int256, mapping, string, uint, uint8, uint16, uint24, uint32, uint40, uint48, uint56, uint64, uint72, uint80, uint88, uint96, uint104, uint112, uint120, uint128, uint136, uint144, uint152, uint160, uint168, uint176, uint184, uint192, uint200, uint208, uint216, uint224, uint232, uint240, uint248, uint256, var, void, ether, finney, szabo, wei, days, hours, minutes, seconds, weeks, years},	% types; money and time units
	keywordstyle=[2]\color{teal}\bfseries,
	keywords=[3]{block, blockhash, coinbase, difficulty, gaslimit, number, timestamp, msg, data, gas, sender, sig, value, now, tx, gasprice, origin},	% environment variables
	keywordstyle=[3]\color{violet}\bfseries,
	identifierstyle=\color{black},
	sensitive=true,
	comment=[l]{//},
	morecomment=[s]{/*}{*/},
	commentstyle=\color{gray}\ttfamily,
	stringstyle=\color{red}\ttfamily,
	morestring=[b]',
	morestring=[b]"
}

\newcommand{\code}[1]{\text{\lstinline[
	language=Rust,
	style=colouredRust,
	basicstyle=\small\ttfamily,
	mathescape=true,
	escapechar=~,
]`#1`}}

\lstdefinelanguage{stdout}
{
	morecomment=[l]{//},
	morecomment=[l]{[...]},
	commentstyle=\color[gray]{0.4},
}

\lstnewenvironment{stdout}{%
	\lstset{
		language=stdout,
		style=boxed,
		basicstyle=\small\ttfamily,
		mathescape=true,
		escapechar=~,
		tabsize=2,
	}
}{}

\def\naive{na\"{\i}ve}

\newcommand{\wrt}{{{w.r.t.\@}}}

\newcommand{\eg}{{{e.g.\@}}}

\newcommand{\ie}{{{i.e.\@}}}
\newcommand{\cf}{{{cf.\@}}} % None of the other abbreviations are \it, so why should this one be
\newcommand{\myedit}[2]{\ifdefined\noeditcolour{#2}\else{\color{#1}{#2}}\fi}
\DeclareRobustCommand{\myout}[2]{\ifdefined\noeditcolour{}\else\myedit{#1}{\texorpdfstring{\sout{#2}}{#2}}\fi}
\newcommand{\myfootnote}[3]{\ifdefined\noeditcolour{}\else\myedit{#2}{\footnote{\myedit{#2}{#1:~ #3}}}\fi}
\def\zgcolour{violet!80}

\def\zgfootnote{\myfootnote{ZG}{\zgcolour}}

\def\zg{\myedit{\zgcolour}}
\def\zgout{\myout{\zgcolour}}
\newcommand{\zgedit}[2]{\zgout{#1 }{\zg{#2}}}

\lstdefinelanguage{sil}{
    morekeywords={
        label,
        result,
        Int, Perm, Bool, Ref,
        true, false,
        null,
        method, function, predicate, domain, axiom, var, returns,
        requires, ensures, invariant,
        fold, unfold, inhale, exhale, new, assert, assume, goto,
        while, if, elseif, else,
        fresh,
        unfolding, in,
        old,
        forall, exists,
        acc, wildcard, write, none, epsilon, perm,
        Seq, Set, MultiSet, union, intersection, setminus, subset
        unique
    },
    sensitive=true,
    morecomment=[l]{//},
    morecomment=[s]{/*}{*/},
    morestring=[b]",
    firstnumber=1,
    numberstyle=\tiny,
    basicstyle=\ttfamily\small,
    keywordstyle=\bfseries,
    ndkeywordstyle=\bfseries,
}
\lstnewenvironment{sil}[1][]{%
    \lstset{language=sil,
        escapechar=~,
        floatplacement={tbp},captionpos=b,
        frame=lines,xleftmargin=8pt,xrightmargin=8pt,#1}}{}

\newcommand{\scite}[1]{\unskip~\cite{#1}}

\newcommand{\drawredsquig}[2]{%
  \begin{tikzpicture}[overlay, remember picture]
    \draw[red, line width=1pt, decorate, decoration={snake, amplitude=0.8pt, segment length=4pt}]
      ($(pic cs:#1)+(0,-2pt)$) -- ($(pic cs:#2)+(0,-2pt)$);
  \end{tikzpicture}%
}

\newcommand{\numtowords}[1]{%
    \ifthenelse{\equal{#1}{1}}{one}{%
    \ifthenelse{\equal{#1}{2}}{two}{%
    \ifthenelse{\equal{#1}{3}}{three}{%
    \ifthenelse{\equal{#1}{4}}{four}{%
    \ifthenelse{\equal{#1}{5}}{five}{%
    \ifthenelse{\equal{#1}{6}}{six}{%
    \ifthenelse{\equal{#1}{7}}{seven}{%
    \ifthenelse{\equal{#1}{8}}{eight}{%
    \ifthenelse{\equal{#1}{9}}{nine}{%
    #1%
    }}}}}}}}}%
}

%% file: statmacros.tex
\newcommand{\comparisonlineratio}{48\%}
\newcommand{\comparisondepthratio}{38\%}
\newcommand{\comparisonsizeratio}{41\%}
\newcommand{\numcomparisons}{31}
\newcommand{\numworsecomparisons}{3}
\newcommand{\numbettersize}{26}
\newcommand{\numbetterdepth}{29}
\newcommand{\numbettercomparisons}{25}
\newcommand{\numsamelinecomparisons}{3}

%% file: abstract.tex
Specifications for modular program verifiers are expressed as constraints on
program states (\eg{} preconditions) and relations on program states (\eg{}
postconditions). For programs whose domain is managing \emph{resources} of any
kind (\eg{} cryptocurrencies), such state-based specifications must make
explicit properties that a human would implicitly understand for free. For
example, it's clear that depositing into your bank account will not change other
balances, but classically this must be stated as a \emph{frame condition}. As a
result, classical specifications for resource-manipulating programs quickly
become verbose and difficult to interpret, write and debug.

In this paper, we present a novel methodology that \zgedit{introduces
user-defined first-class resources in the specification language}{extends a
modular program verifier to support \emph{user-defined first-class resources}},
allowing resource-related operations and properties to be expressed directly and
eliminating the need to reify implicit knowledge in the specifications. We
implement our methodology as an extension of the program verifier Prusti, and
use it to verify \zg{real-world smart contracts and} a key part of a
\zgout{real-world} blockchain application. \zgedit{As we demonstrate in our
evaluation,}{Our evaluation demonstrates that} specifications written with our
methodology are more concise\zgedit{, syntactically simpler, and easier to
understand}{ and substantially simpler} than \zgout{alternative}specifications written
purely in terms of program states.

%% file: introduction.tex
\section{Introduction}

The goal of program verification is to ensure that a program will always operate
in accordance with a formal specification. In practice, this requires encoding
the desired properties in the specification language of a program verifier that
will be used to check conformance to such properties. The difference between our
intuitive, human language description of desired properties and their
representation in the specification language is called the \emph{semantic
gap}~\cite{semanticgap}. The larger this gap is in practice, the more difficult
it becomes to write, read, maintain, and debug specifications.

In modular program verifiers, specifications are written and checked for each
function separately\zg{, with the guarantee that separately verified functions
compose to a verified whole}\zgfootnote{Requested by reviewer B}. Specification
preconditions describe requirements on the program state at function call sites,
and postconditions describe the corresponding guarantees about the program state
at return sites; the verifier \zg{checks that, when the precondition holds, the
function's body ensures the postcondition, and}\zgfootnote{Requested by reviewer
B} reasons about function calls with respect to these specifications rather than
the function's implementation. For example, the specification for a function
\texttt{sort()} that sorts a list in-place could be written as an expression
relating the values of the input list both before and after the call; namely,
stating that the latter should be a sorted \zgedit{version}{permutation} of the
former. This design effectively minimises the semantic gap for properties
naturally expressed as relations on program states, such as sortedness.

However, properties about programs that manipulate resources, such as money, are
more challenging to specify correctly in this manner. For example, in a banking
application, one might consider describing the effect of a function
\texttt{deposit(acct\_id, amt)}, using a postcondition:
\begin{rust}
#[ensures(bank.balance(acct_id) == old(bank.balance(acct_id)) + amt)]
\end{rust}
which specifies that after calling \texttt{\zg{bank.}deposit()} the balance for
\texttt{acct\_id} will have increased \zgedit{appropriately}{by \texttt{amt}}.
However, this specification misses an important property: the balance of all
\emph{other} accounts in the bank must remain \emph{unchanged} after the
call\zg{\footnote{\zg{Strictly speaking, all other aspects of program state
(\eg{} objects other than \texttt{bank}) should also remain unchanged, however
many program verifiers provide some coarse-grained mechanism to facilitate such
framing.}}}. To address this, the traditional solution
is to extend the specification with a \emph{frame
  condition}\scite{frameproblem}: an assertion that besides the change to
\texttt{balance(acct\_id)}, everything \emph{else} stays the same.
This results in the postcondition:
\begin{rust}
#[ensures(bank.balance(acct_id) == old(bank.balance(acct_id)) + amt &&
  forall(|a_id: AcctId| a_id != acct_id ==>
    bank.balance(a_id) == old(bank.balance(a_id))))]
\end{rust}

This specification is longer and more complicated; most of the complexity stems
from describing what's \emph{unchanged} by \texttt{deposit()} (this extra
complexity grows rapidly for more complicated functions). But in terms of an
intuitive understanding of bank deposits, this property simply
\zgedit{doesn't}{does not} \emph{need} to be stated: it's obvious to humans that
depositing money into one account will not change other balances. This suggests
a wide semantic gap between the way we think about resources in the real-world,
and what can be expressed in classical techniques for formal specification.

An elegant way to close this gap is to introduce resources into the
specification language itself. Various powerful frameworks (\eg{}
Iris\scite{iris}) facilitate this via substructural logic (\eg{} separation
logic\scite{reynolds2002separation}).
However, many modern, widely-used program verifiers such as
Boogie\scite{boogie}, Dafny\scite{Dafny}, and Why3\scite{why3} use classical
logic.
Although, in general, such verifiers provide framing guarantees on a
per-data-structure level (\eg{} the \texttt{modifies} clauses in Boogie), we
observe that for fine-grained changes there is still a need to define frame
conditions via pointwise assertions on program state, even in state-of-the art
program verifiers.

In this paper,\zgout{we present a novel specification methodology that narrows
this gap by introducing custom resources and a language of resource properties
and operations as \emph{first-class elements} of our specification language}
\zg{we present a novel methodology for extending an existing specification
language with custom resources and operations in a lightweight manner}. Our
\zgedit{methodology}{approach} allows resource-related properties to be
expressed directly, \zgedit{and eliminates}{eliminating} the need to reify
properties that are implicit in our intuitive understanding of resources, while
building in ubiquitous properties by default, \eg{} that the amount of a
resource should remain constant unless explicitly created or destroyed.

\zgedit{The usage of our custom resources in specification methodology is an
abstraction}{In our methodology, resources used in specifications are
abstractions} decoupled from the \zgedit{underlying program representations of
these notions}{program state}. However, our technique allows \textit{coupling
invariants} to be defined, specifying once and for all the mapping between
abstract resource notions and concrete program states. \zg{In contrast to more
sophisticated methodologies (\eg{} for fine-grained concurrency), our coupling
invariants are not required to be reflexive or transitive, and can be desugared
via a straightforward syntactic translation.} Specifications per function can be
written concisely and more simply using resource abstractions, and the coupling
invariants automatically and implicitly entail the right proof obligations
concerning changes to and constraints on concrete program states.

Our paper makes the following main contributions:

\begin{enumerate}
    \item We demonstrate the typical problems that arise when writing specifications of
          resource-manipulating programs in terms of program states (\secref{motivation})
    \item We present a methodology to solve these problems using
          first-class resources (\secref{design})
    \item We implement our methodology, extending the Prusti verifier (\secref{implementation})
    \item As a case study, we apply our methodology and implementation to verify a key component of a real-world Rust implementation of a
        cross-chain token transfer protocol (\secref{casestudy})
    \item We perform a comparative evaluation of our methodology, showing that
it enables\zgout{users to write} shorter and simpler specifications for
\zg{real-world} resource-manipulating programs (\secref{evaluation})
\end{enumerate}
%We discuss related work in \secref{related}, concluding in \secref{conclusion}. We begin with a more in-depth motivation.

%% file: motivation.tex
\section{Motivation}\label{sec:motivation}

\begin{figure}[h]
\begin{rust}
type AcctId = u32;
trait Bank {
  fn balance(&self, acct_id: AcctId) -> u128;
  fn deposit(&mut self, acct_id: AcctId, amt: u128);
  fn withdraw(&mut self, acct_id: AcctId, amt: u128);
}
\end{rust}
\caption{\zgedit{Rust code defining a \texttt{Bank} struct and implementation of
banking operations. Account balances are stored in a \texttt{U32Map} mapping
\texttt{AcctId}s to \texttt{u32} values; deposits and withdrawals update the map
entries.}{Rust code defining a \texttt{Bank} interface. Monetary values are
represented as unsigned 128-bit integers, account identifiers are represented by values
of type \texttt{AcctId} (\texttt{u32}).}}\label{fig:motivation:bankinterface}
\end{figure}

To motivate our methodology, we first present the challenges that arise when
writing specifications for a resource-manipulating program. Throughout this
section, we consider a Rust \zgedit{implementation of}{interface for} a
multi-account bank that stores (dollar) balances, shown in
\figref{motivation:bankinterface}.
\zgedit{
The struct \texttt{Bank} manages the balances for multiple accounts in the field
\texttt{map}, each account is identified by a corresponding \texttt{AcctId}.
Clients of the \texttt{Bank} cannot access \texttt{map} directly; they
interact with the \texttt{Bank} object via the functions \texttt{balance},
\texttt{deposit}, and \texttt{withdraw}, within the \texttt{impl Bank
\{...\}} block.
}{
Clients interact with a \texttt{Bank} via the functions \texttt{balance},
\texttt{deposit}, and \texttt{withdraw}.
}

These functions take as first parameter a reference to \zgedit{the}{a}
\texttt{Bank} instance. The syntax \texttt{\&self} indicates a Rust \emph{shared
reference}, with which the \texttt{balance} function cannot modify the
\texttt{Bank}, while a \emph{mutable reference} such as the \texttt{\&mut self}
parameter of the other functions allows modifications.
\zgout{A \texttt{Bank} stores account balances in a \texttt{U32Map} that maps
account identifiers to \texttt{u32} values. The \texttt{balance()} function
looks up an account identifier in the map, returning \texttt{0} if there is no
entry. The functions \texttt{deposit()} and \texttt{withdraw()} update the map,
overwriting any existing entry.}

Specifications for the \texttt{Bank} \zgedit{API}{interface} play two main
roles: defining what it means for \zgedit{the}{an} \emph{implementation} of
these functions to be correct (\eg{} \zg{an implementation of} \texttt{withdraw}
must correctly update the \zgedit{right}{correct} balance \zg{internally}), and
defining what \emph{client code} should consider when using this \zgedit{API}{interface}. For
example, withdrawing more than the current balance should be forbidden, while to
specify (and modularly verify) a client function such as \code{transfer} below
relies on having suitable specifications for \texttt{withdraw} and
\texttt{deposit}:

\begin{rust}
fn transfer<B: Bank>(bank: &mut B, from: AcctId, to: AcctId, amt: u128) {
  bank.withdraw(from, amt); bank.deposit(to, amt);
}
\end{rust}

\zgout{Our bank example is substantially less complex than typical real-world
code but}\zg{Real-world programs are substantially more complex than our bank
example, although we note that the \texttt{Bank} trait itself is derived from
similar interfaces found in real-world programs (\eg{} the \texttt{BankKeeper}
trait described in our case study [\secref{casestudy}]). Nonetheless, even for
our simplified example, }writing correct specifications\zgout{for it} in a classical
fashion\zg{, \ie{} via pointwise assertions on the domain of the \texttt{balance}
function,} can nonetheless be challenging and subtle. We now demonstrate three
specific issues that we encounter when trying to specify this kind of code.

\begin{figure}[h]
\begin{rust}
#[ensures(
\end{rust}\vspace{-4mm}
\begin{rustbad}[2]
  forall(|acct_id2: AcctId| if acct_id == acct_id2 {
\end{rustbad}\vspace{-4mm}
\begin{rust}[3]
    self.balance(acct_id) == old(self.balance(acct_id)) + amt
\end{rust}\vspace{-4mm}
\begin{rustbad}[4]
  } else {
    self.balance(acct_id2) == old(self.balance(acct_id2))
  })
\end{rustbad}\vspace{-4mm}
\begin{rust}[7]
)]
fn deposit(&mut self, acct_id: AcctId, amt: u128);

#[requires(self.balance(acct_id) >= amt)]
#[ensures(
\end{rust}\vspace{-4mm}
\begin{rustbad}[12]
  forall(|acct_id2: AcctId| if acct_id == acct_id2 {
\end{rustbad}\vspace{-4mm}
\begin{rust}[13]
    self.balance(acct_id) == old(self.balance(acct_id)) - amt
\end{rust}\vspace{-4mm}
\begin{rustbad}[14]
  } else {
    self.balance(acct_id2) == old(self.balance(acct_id2))
  })
\end{rustbad}\vspace{-4mm}
\begin{rust}[17]
)]
fn withdraw(&mut self, acct_id: AcctId, amt: u128);
\end{rust}
\caption[Specifications for the \texttt{deposit} and \texttt{withdraw}
functions]{Specifications for the \texttt{deposit} and \texttt{withdraw}
functions. Each includes an explicit frame condition;
specification lines used to encode the frame condition are highlighted in red.}\label{fig:motivation:bankspec}
\end{figure}

\subsection{Three \Issues{} of Specifying Resource-Manipulating
Programs}\label{sec:motivation:noresources}

A typical specification for our \texttt{Bank} \zgedit{implementation}{interface}
would define postconditions for \texttt{deposit} and \texttt{withdraw} in
terms of calls to \texttt{balance}. As discussed in the introduction, an
intuitive (but ultimately insufficient) postcondition for \texttt{deposit}
could be written as follows:
%\begin{figure}[t]
\begin{rust}
#[ensures(self.balance(acct_id) == old(self.balance(acct_id) + amt))]
fn deposit(&mut self, acct_id: AcctId, amt: u128)
\end{rust}

Although \zgedit{this is a correct postcondition for the implementation}{a
correct implementation should satisfy this postcondition}, it is not sufficient
for (modular) client reasoning: it specifies the change to
\texttt{balance(acct\_id)} but doesn't say anything about \emph{other} balances
(which should be unchanged). The standard solution is to extend the
postcondition with a \emph{frame condition}, specifying that the balances of all
other accounts remain unchanged. The \texttt{withdraw} function also requires
similar frame conditions. A correct such specification for \texttt{deposit},
and the analogous specification for \texttt{withdraw}, are presented in
\figref{motivation:bankspec}. Adding frame conditions makes the resulting
specifications more complex: much of the specification effort is concerned with
precisely specifying what does \emph{not} change. Similar frame conditions are
also necessary for any function with a mutable reference to \zgedit{the}{a}
\texttt{Bank}, such as \texttt{transfer} above, whose postcondition must
specify overall what has not changed.

Deriving a frame condition for \texttt{transfer} is not simply a matter of,
say, conjoining those for the two called functions; instead, one should rethink
these specifications as describing \emph{state updates}, mentally compose these
updates, and deduce the right end-to-end specification of their composition: the
appropriate postcondition of \texttt{transfer} is presented in
\figref{motivation:transferspec}. Note that the postcondition does not directly
use the specification expressions from \texttt{deposit} or
\texttt{withdraw}.

\begin{figure}[h]
\begin{rust}
#[requires(from != to && bank.balance(from) >= amt)]
#[ensures(forall(|a_id: AcctId|
  if      a_id == from { bank.balance(a_id) == old(bank.balance(a_id)) - amt }
  else if a_id == to   { bank.balance(a_id) == old(bank.balance(a_id)) + amt }
  else                 { bank.balance(a_id) == old(bank.balance(a_id))       }))]
fn transfer<B: Bank>(bank: &mut B, from: AcctId, to: AcctId, amt: u128) { ... }
\end{rust}
\caption[The specification for the \texttt{transfer()} function]{A specification for the \texttt{transfer} function, which makes
  calls to \texttt{withdraw} and then \texttt{deposit}
}\label{fig:motivation:transferspec}
\end{figure}

This demonstrates the first
\issue{} with classical specifications in this context:

\callout{\Issue{} 1:}{Specifying resource operations using program states
  requires frame conditions}

Deriving frame conditions for the composition of multiple operations requires
syntactic and mental overhead. But similar composition issues can arise even
without consideration of frame conditions. For example, consider the function
\texttt{withdraw2} in \figref{motivation:withdraw2}, which performs two
withdrawal operations in sequence. When we consider how to specify the
precondition of this function, based on the precondition of the
\texttt{withdraw} function (\figref{motivation:bankspec}), we might consider
using:
\begin{rust}
#[requires(bank.balance(acct_id1) >= 1 && bank.balance(acct_id2) >= 2)]
\end{rust}

This precondition is not strong enough however: when \texttt{acct\_id1 =
acct\_id2}, the \texttt{withdraw2} function \zgedit{would withdraw}{withdraws}
\$3 from the account. Therefore, a precise precondition needs to explicitly
branch on whether \zgedit{both identifiers are the same}{the identifiers are
equal}. Furthermore, the analogous issue exists with the \emph{postcondition}:
the way we represent the change in program state differs depending on
\zgedit{the aliasing of the parameters}{whether the parameters are equal}. An
accurate specification for the \texttt{withdraw2} function is presented in
\figref{verifying:withdraw2spec}.

\begin{figure}[h]
\begin{rust}
fn withdraw2<B: Bank>(bank: &mut B, acct_id1: AcctId, acct_id2: AcctId) {
  bank.withdraw(acct_id1, 1); bank.withdraw(acct_id2, 2);
}
\end{rust}
\caption{A function that performs two withdrawals in sequence. When
\texttt{acct\_id1} and \texttt{acct\_id2} are the same, this function performs
two operations on the same account.}\label{fig:motivation:withdraw2}
\end{figure}

\begin{figure}[h]
\begin{rust}
#[requires(if(acct_id1 == acct_id2) { bank.balance(acct_id1) >= 3 }
  else { bank.balance(acct_id1) >= 1 && bank.balance(acct_id2) >= 2 })]
#[ensures(if(acct_id1 == acct_id2) {
  bank.balance(acct_id1) == old(bank.balance(acct_id1) - 3)
} else {
  bank.balance(acct_id1) == old(bank.balance(acct_id1) - 1) &&
  bank.balance(acct_id2) == old(bank.balance(acct_id2) - 2)
} && forall(|acct_id: AcctId| (acct_id != acct_id1 && acct_id != acct_id2) ==>
    bank.balance(acct_id) == old(bank.balance(acct_id))
))]
fn withdraw2<B: Bank>(bank: &mut B, acct_id1: AcctId, acct_id2: AcctId) { ... }
\end{rust}
\caption{The specifications for the \texttt{withdraw2} function in
\figref{motivation:withdraw2}. The conditional in the precondition is necessary
to handle the possible aliasing between \texttt{acct\_id1} and
\texttt{acct\_id2}. The postcondition includes an analogous conditional, as well
as a frame condition stating that other account balances remain unchanged.  }\label{fig:verifying:withdraw2spec}
\end{figure}
Again, to systematically derive such a specification requires considering the
\emph{updates} described by the specifications of the called functions, mentally
composing these side-effects, and mapping the resulting transformation back to a
state-based specification, considering separate cases corresponding to the
possible (in-)equalities between parameters as needed.
The complexity of this process increases significantly as the number of
operations increases. For example, the specification for a hypothetical
\texttt{withdraw3} function would need to branch on all
\zgedit{aliasing}{equality} possibilities between three parameters. This
illustrates the second issue:

\callout{\Issue{} 2:}{Specifications of resource operations using program states
do not easily compose}

The specifications for the \texttt{Bank} should dictate how clients are allowed
to interact with it. They prevent \eg{} overdrawing an account via the
precondition \texttt{self.balance(acct\_id) >= amt} on \texttt{withdraw}. But
they do not enforce \emph{inherent} properties of how resources should behave.
Note that \texttt{deposit} does not have a precondition: the specifications do
not restrict calls to \texttt{deposit}.

\begin{figure}[t]
\begin{rust}
#[requires(bank.balance(from) >= amt)]
fn transfer<B: Bank>(bank: &mut B, from: AcctId, to: AcctId, amt: u128) {
  bank.withdraw(from, amt); bank.deposit(to, amt); bank.deposit(to, amt);
}
\end{rust}
\caption{A buggy transfer function that performs two deposits instead of one.
\zg{It will be accepted by the verifier, despite creating money out of
nowhere.}}\label{fig:motivation:badtransfer}
\end{figure}

\figref{motivation:badtransfer} shows an alternative (faulty) \texttt{transfer} implementation:
it deposits more money in account \texttt{to} than was withdrawn from account
\texttt{from}, effectively creating money out of nowhere. The specifications \zgedit{of
the}{for} \texttt{Bank} do not prevent this behaviour: the \texttt{transfer}
function verifies. Technically, \texttt{transfer} is underspecified: with the
complete functional specification of its intended effect on the \texttt{to} account, the function \emph{would} fail to verify. But the conceptual source of the bug is clear from the code even with a weak specification: it's wrong for a resource acquired once to be spent twice. Intuitively, the money spent by \texttt{deposit} should come from somewhere, but this relies on an understanding of the specific resources the program works with, and of properties \emph{all} resources should have, that is only partially explicit in the program state and its classical specifications.

% However, ideally,
%such a function shouldn't verify in the first place, because the second
%\texttt{deposit()} operation cannot be justified.

%The root problem is that the \texttt{Bank} specifications aren't sufficient to
%ensure that calls to \texttt{deposit()} or \texttt{withdraw()} are permissible;
%i.e., that they could would be allowed by a real bank handling physical money.
%Intuitively, the money from a \texttt{deposit()} needs to have come from
%somewhere, yet the \texttt{Bank} does not impose a corresponding precondition. 
%
%A
%similar argument also applies to \texttt{withdraw()}: being able to prove that
%an account has a certain amount of funds is not the same as being allowed to
%withdraw them.

\callout{\Issue{} 3:}{Specifications using program states cannot easily enforce
  resource properties}

The root cause underlying all three of these issues is that program states are
not the ideal abstraction for expressing and reasoning about the behaviour of
resource-manipulating programs. Intuitively, we think about such programs in
terms of their operations on resources (\eg{} spending money),
\zgedit{including}{incorporating} implicit common-sense understanding of how
resources work in reality: for example, resources cannot be duplicated. The key
idea behind our work is to make this abstraction directly available in the
specification language, reifying this informal intuition and all its benefits in
a formal technique. Our technique is presented concretely in the next section,
but the basic idea is to allow (possibly multiple) custom \rtypes{} to be
declared for a given program\zgout{: we might use a \rtype{} to represent money known
to be stored in the bank and/or a \rtype{} to represent money taken \emph{out}
of the bank}. The notion of \emph{how much} of each of these resources is
currently held is made a (ghost) part of the program state (the \emph{resource
state}), and new specification features are provided that define how much
resource is \emph{transferred} in and out of function calls.

The default interpretation of these transfers is that if we don't specify that one happens, it \emph{cannot}, allowing specifications to describe local updates without explicit frame conditions (\Issue{} 1). Since resource transfers directly express \emph{changes} to the resource state, they compose easily (\Issue{} 2). Using a custom resource to represent \emph{withdrawn} money prevents the faulty code in \figref{motivation:badtransfer}, since the attempt to spend a resource which is no-longer held can be made explicit with our first-class resources and the built-in properties that come with our resource model (\Issue{} 3).

In the next section, we demonstrate how our methodology enables this kind of reasoning.

%% file: methodology.tex
\section{Methodology}\label{sec:design}\label{sec:methodology} % aliasing ftw

The core idea of our approach is to enable program behaviour to be specified
directly in terms of notions of resource relevant to the program.
\emph{Resources} in our methodology are conceptually a pair of a \emph{\rtype{}}
and a \emph{resource amount} (an integer). Resource types are organised into
parameterised \emph{\rkind}s: a named \emph{\rtypecon} with fixed arity
(possibly zero) and corresponding parameter types; each instantiation of these
parameters yields a distinct \emph{\rtype{}}. \zg{Any value that can be checked
for equality and represented in the specification syntax can be used as a
parameter to a \rtypecon{}.} For example, we might use a \rkind{} \texttt{Money}
with a single parameter \zgedit{representing an account ID}{of type
\texttt{AcctId}} to represent currency known to be deposited in the
corresponding bank account.

In our Prusti implementation, such a \rkind{} is declared as a Rust struct annotated with
a \resourceann{} tag, listing its parameter types; \rtype{}s are denoted by instantiating the struct with
appropriate (Rust) expressions. For example, we can declare a \rtypecon{} for the
amount of money belonging to a particular account as follows:
\begin{rust}
~\resourceann{}~ struct Money(AcctId);
\end{rust}
Resources of the same \rtypeshort{} are always aggregated to the sum of their
amounts, and resources of different \rtypeshort{}s are incomparable. In our
example, the fact that for two \emph{different} account IDs \texttt{id1} and
\texttt{id2}, the corresponding resource types \texttt{Money(id1)} and
\texttt{Money(id2)} are different encodes the property that money in different
accounts should not be fungible; withdrawing money from your account should have
a meaning distinct from withdrawing from someone else's!

We use the syntax \transferskwd\texttt{(}\textit{rtype}, \textit{amt}\texttt{)} to denote resources, where \textit{rtype} is a \rtype{}, and \textit{amt} is the
amount. For example, the expression \texttt{\transfers{Money(a\_id), amt}}
refers to an amount \texttt{amt} of money for the account associated with the
identifier \texttt{a\_id}.

\subsection{Resource Operations}\label{sec:design:ops}

\begin{figure}[h]
\begin{rust}
struct BankImpl { map: HashMap<AcctId, u128> };
impl Bank for BankImpl {
  fn balance(&self, acct_id: AcctId) -> u128 {
    self.map.get(acct_id).unwrap_or(0)  // 0 returned if no entry in map
  }
  fn deposit(&mut self, acct_id: AcctId, amt: u128) {
    let bal = self.balance(acct_id); self.map.insert(acct_id, bal + amt);
    ~\inhalekeyword(\transfers{Money(acct\_id), amt}~);
  }
  fn withdraw(&mut self, acct_id: AcctId, amt: u128) {
    let bal = self.balance(acct_id); self.map.insert(acct_id, bal - amt);
    ~\exhalekeyword(\transfers{Money(acct\_id), amt}~);
  }
}
\end{rust}
\caption{Rust code defining a \texttt{BankImpl} struct and implementation of banking
operations. Account balances are stored in a
\zgedit{\texttt{U32Map}}{\texttt{HashMap}} mapping \texttt{AcctId}s to
\texttt{\zgedit{u32}{u128}} values; deposits and withdrawals update the map
entries. The \inhalekeyword{} statement in \texttt{deposit} denotes the creation of the
resource representing \texttt{amt} units of money for account \texttt{acct\_id},
the \exhalekeyword{} statement in \texttt{withdraw} denotes its
destruction.}\label{fig:methodology:bankimpl}
\end{figure}

\zg{Conceptually, each stack frame owns a collection of resources}. Our
methodology defines three \emph{resource operations}: resources can be
\textit{created} \zg{(added to a stack frame)}, \emph{destroyed} \zg{(removed
from a stack frame)}, and \emph{transferred} \zg{(between stack frames, at call
and return sites)}. We denote creation and destruction by \inhalekeyword{} and
\exhalekeyword{} respectively\zg{, they would appear \eg{} in an
\emph{implementation} of the \texttt{Bank} trait
(\figref{methodology:bankimpl}). In our \texttt{Bank} example, using the
resource type \texttt{Money(acct\_id)} to represent deposited currency, deposits
create this resource and withdrawals destroy it.} Intuitively, it is only
possible to destroy a resource if it is currently held: our methodology requires
that the resource \transfers{Money(acct\_id), amt} is available to destroy at
the point of the \exhalekeyword{} inside \texttt{withdraw} (otherwise verification
will fail).

\begin{figure}[h]
\begin{rust}
trait Bank {
  fn balance(&self, acct_id: AcctId) -> u128;

  #[ensures(~\transfers{Money(acct\_id), amt}~)]
  fn deposit(&mut self, acct_id: AcctId, amt: u128);

  #[requires(~\transfers{Money(acct\_id), amt}~)]
  fn withdraw(&mut self, acct_id: AcctId, amt: u128);
}
\end{rust}
\caption{The specification of \texttt{Bank}, described in terms of the resource
\texttt{Money}.}\label{fig:design:rspec}
\end{figure}

\zg{Each function call performs a transfer of resources based on the
\zgedit{function's}{callee's} specification: conceptually, the resources in a
\zgedit{function's}{callee's} precondition are transferred to it from its
caller, and it returns to the caller the resources in its postcondition. The
\texttt{Bank} trait would be specified as in \figref{design:rspec}:
\texttt{Bank.deposit} returns an amount \texttt{amt} of the resource
\texttt{Money(acct\_id)} to the caller, \texttt{Bank.withdraw} requires the
caller to provide it.}
\zgout{In our \texttt{Bank} example, using the resource
type \texttt{Money(a\_id)} to represent deposited currency, deposits should
create resource and withdrawals destroy it. This idea is realised by
instrumenting the bodies of the \texttt{deposit} and \texttt{withdraw} functions
as shown in \figref{design:rspec}.}

\zgout{ Conceptually, the caller of \texttt{withdraw} needs to give up this
resource: we express that they must \emph{transfer} it from their (calling)
context by specifying the resource in the function precondition (\cf{}
\figref{design:rspec}). Analogously, we specify that \texttt{deposit()}
transfers a resource back to the caller by placing it in the function's
postcondition.
}

\subsection{Resource
State}\label{sec:design:resourcestate}\label{sec:design:resourcetransfer} Our
methodology associates every stack frame with a \emph{resource state}, tracking
the resources currently held by the function invocation. This is \emph{ghost
state}; we use it as a concept for static reasoning about the program only, but
it need not (and will not) actually exist at runtime. A resource state is a map
from \rtype{}s to resource amounts, and resource operations are interpreted with
respect to the current resource state. Creation of a resource
\zgedit{increments}{increases} the corresponding map entry, while destruction
\zgedit{decrements}{decreases} it. Destruction operations entail a proof
obligation: we need to prove that sufficient resource is held before it is
destroyed, otherwise a verification error is raised.%
%
%In our methodology, resource operations are interpreted with respect to their
%effects on the \emph{resource state}: a symbolic map from \rtype{}s to resource
%amounts. The resource state represents the resources available at a given point
%in the program.
%
%In our methodology, the verifier interprets a creation operation as an action
%that increases the amount of the resource in the resource state for the
%corresponding resource type; destruction operations have the opposite effect.

%\subsubsection{Resource Transfers}\label{sec:design:resourcetransfer}
%
Resources are transferred between stack frames at call (and return) sites. The
resource state of each frame is initialised with \emph{only} the resources
transferred to it from its caller, according to the \zgedit{function}{callee}'s
precondition. Resources in the \zgedit{function}{callee}'s postcondition are
those it transfers back to its caller. Transfers also entail proof obligations:
the verifier must prove that the resources transferred away are actually held.
This model of resource state is suitable for modular verification: each
function's initial resource state is determined purely by its precondition, independently of call site.
%
%In our design, resources in the precondition describe the initial resource state
%of the function (which are transferred into the stack frame during calls to the
%function), and resources in the postcondition are those that the function
%returns to its caller (and are checked to be present in the callee's resource state at the
%end of the function).

\subsection{\RSE{}s}\label{sec:design:holdfunc}

The specifications presented in \figref{design:rspec} suffice to characterise
the behaviour of the program in terms of its resource operations. Reifying these
resource transfers \zgedit{make}{makes} some important properties of the program
clearer: \eg{} it is impossible to overdraw an account because withdrawals
require a resource representing at least part of the current balance (see
\figref{design:withdraw2}).

\begin{figure}[h]
\begin{rust}
fn bad<B: Bank>(bank: &mut B, acct_id: AcctId, amt: u128) {
  bank.deposit(acct_id, amt); bank.withdraw(acct_id, amt);
  ~\tikzmark{design:vfstart}~bank.withdraw(acct_id, amt)~\tikzmark{design:vfend}~; // ERROR: insufficient permission to resource
}
\end{rust}
\drawredsquig{design:vfstart}{design:vfend}
\caption{An example program that would not verify after annotating \texttt{Bank}
with resource operations. The first \texttt{withdraw()} is permitted, because
sufficient resource \texttt{Money(acct\_id)} is made available by the preceding
\texttt{deposit()} call. A verification error is raised because there is no
resource available for the second \texttt{withdraw()}.}\label{fig:design:withdraw2}
\end{figure}

However, our resource specifications so far are completely independent of the
\emph{actual} \zgedit{implementation}{program} state used to track resources. In
particular, there is no specified relationship between resource operations and
\texttt{Bank.balance()} calls. Working towards this, we introduce
\emph{\rse{}s}, which are used inside specifications to introspect on the
resource state. These expressions are written \holds{\texttt{rtype}}, and (when
used \emph{outside} of function specifications \eg{} as inline assertions)
denote the amount of resource type \texttt{rtype} held in the current resource
state. For example, \holds{\texttt{Money(acct\_id)}} denotes the amount of
\texttt{acct\_id}'s money in the current state.

When \holdskeyword{} is used in function specifications, we define its semantics
differently, considering the fact that the caller and callee resource states may
differ. In such positions, our technique interprets \holds{} expressions with
respect to the \emph{amount of resource transferred so far} by the corresponding
pre-/post-condition; this notion has the same meaning for both caller and
callee.

\figref{design:resourcespec} illustrates this semantics.
The first precondition \texttt{\#[requires(resource(Money(a), 1))]} at line
1 specifies a resource to be transferred
from caller to callee. As that is the only resource transferred so far, the
(commented) assertion \texttt{holds(Money(a)) == 1} holds at \Circled{1}. The
subsequent line specifies the same resource transfer again, resulting in 
\texttt{holds(Money(a)) == 2} (\Circled{2}).

The first
postcondition \texttt{old(holds(Money(a))) == 2)} at \Circled{3} concerns (via \texttt{old}) the 
amount of \texttt{Money(a)} transferred \emph{from the caller} (this refers to the overall transfers made by the preconditions together), while the second (\Circled{4} ) states that no money has been transferred by the \emph{postconditions} up to this point.
At \Circled{5}, located after the postcondition transferring \texttt{resource(Money(a),1)} to the caller, the commented assertion \texttt{holds(Money(a)) == 1} would instead be true.

\begin{figure}[t]
\begin{rust}
#[requires(resource(Money(a), 1))] // holds(Money(a)) == 1 ~\Circled{1}~
#[requires(resource(Money(a), 1))] // holds(Money(a)) == 2 ~\Circled{2}~
#[ensures(old(holds(Money(a))) == 2)]                      ~\Circled{3}~
#[ensures(holds(Money(a)) == 0)]                           ~\Circled{4}~
#[ensures(resource(Money(a), 1))] // holds(Money(a)) == 1  ~\Circled{5}~
fn take2return1<B: Bank>(bank: &mut B, a:AcctId){ bank.withdraw(a, 1); }

#[requires(resource(Money(a),3))]
fn client<B: Bank>(bank: &mut B, a: AcctId){ take2return1(bank, a); }
\end{rust}
  \caption{A program demonstrating resource operations and \rse{}s in
specifications. \zg{The function \texttt{take2return1} requires 2 units of
  \texttt{Money(a)}} from the caller and returns 1 unit.}\label{fig:design:resourcespec}
\end{figure}

\subsection{Coupling Invariants}

Finally, we use our resource state expressions to define connections between the
resource state and the actual program state. In our methodology, users declare
\emph{coupling invariants} to define these connections. Coupling invariants are
\zgout{expressed as}two-state expressions (using \texttt{old} to describe the prior
state, as in a postcondition) that relate the program \emph{and resource} states
at two points in the program; these invariants are enforced at function return
sites. Coupling invariants are expressed using annotations of the form
\texttt{\#[invariant\_twostate($t$)]}, where $t$ is an expression that can
include specification constructs such as \holds{} and \texttt{old()}
expressions. Although declared once per type, a coupling invariant $t$ is
interpreted as an additional, implicit postcondition on \emph{every} function
taking a mutable reference to this type as a parameter; effectively, it is
equivalent to adding the postcondition \texttt{\#[ensures($t$)]} to all such
functions (this happens implicitly).

\zg{The expressive power of coupling invariants stems from their ability to
refer to the \emph{change} in resource state that occurs within a function (\ie, the
difference between the resources it produces and those it consumes), and relate
it to a corresponding change in program state.
While coupling invariants only consider the function's local resource state,
with program state notionally representing the global amount of resources, we
observe that the \emph{change} in the amount of a resource is the same
regardless of whether we consider all of the resources in a program, or just
those available locally.
Exploiting this property, coupling invariants enable assertions on program
states to be derived from (the aggregate of) a function's resource operations.
}

\begin{figure}[t]
\begin{rust}
#[invariant_twostate(forall(|a_id: AcctId|
  holds(Money(a_id)) - old(holds(Money(a_id))) ==
  self.balance(a_id) - old(self.balance(a_id))))]
\end{rust}
\caption{A coupling invariant for the \texttt{Bank} struct, which is enforced between the beginning (via \texttt{old}) and end of each function involving a mutable reference to a \texttt{Bank}.}\label{fig:design:twostate}
\end{figure}

\begin{figure}[t]
\footnotesize
  \begin{tikzpicture}[
    component/.style={rectangle split, rectangle split parts=2, draw, text centered},
    arrow/.style={-Stealth, line width=0.6pt, dashed, gray}
]

\pgfmathsetmacro{\boxspc}{0.5} % Spacing between boxes
% Spec
\node[component] (spec) {
    \textbf{\texttt{deposit()} Specification (\figref{design:rspec})}
    \nodepart{second}
    \begin{tabular}{ll}
        Pre: & \texttt{<none>} \\
        Post: & \texttt{\#[ensures(\transfers{Money(acct\_id), amt})]}
    \end{tabular}
};

% Resource state
\node[component, below=\boxspc of spec] (resource) {
    \textbf{Resource Transfers Denoted by Specification}
    \nodepart{second}
    \begin{tabular}{ll}
        Transferred In: & $\epsilon$ \\
        Transferred Out: & $\{\texttt{Money}(\texttt{acct\_id}) \to \texttt{amt}\}$
    \end{tabular}
};

% Introspection of Resource state via holds
\node[component, below=\boxspc of resource] (introspection) {
    \textbf{View of Resource Transfers via \texttt{holds()}}
    \nodepart{second}
    \begin{tabular}{ll}
        Before: & $\forall \texttt{a\_id}~.~
        \texttt{old(holds(Money(a\_id)))} = 0$ \\
        After: & $\forall \texttt{a\_id}~.~ \texttt{holds(Money(a\_id))} =
        \begin{cases}
        \texttt{amt} & \text{if } \texttt{a\_id} = \texttt{acct\_id}\\
        0 &  \text{otherwise}
        \end{cases}$
    \end{tabular}
};

% Effect of two_state invariant
\node[component, below=\boxspc of introspection.south east, anchor=north east] (postcondition) {
    \textbf{Automatically Derived Postcondition for \texttt{deposit()} (c.f. \figref{motivation:bankspec})}
    \nodepart{second}
    $\forall \texttt{a\_id}~.~\texttt{balance(a\_id)} = \texttt{old(balance(a\_id))} +
         \begin{cases}
         \texttt{amt} & \text{if } \texttt{a\_id} = \texttt{acct\_id}\\
         0 & \text{otherwise}
         \end{cases}$
};

% invariant
\node[component, left=0.8cm of spec.north west, anchor=north east] (invariant) {
    \textbf{Coupling Invariant}
    \nodepart{second}
    Shown in \figref{design:twostate}
};

% Dashed arrows
\draw [arrow] (spec) -- (resource);
\draw [arrow] (resource) -- (introspection);
\draw [arrow] (introspection.south) -- ([xshift=0.3cm]postcondition.north);
\draw [arrow] (invariant) |- ([yshift=-0.25cm]postcondition.west);

\end{tikzpicture}
\caption{A diagram indicating how the coupling invariant from
\figref{design:twostate} combines with resource operations of
\texttt{deposit()} (\figref{design:rspec}) to derive a postcondition specifying
the effect \texttt{deposit()} has on the value of
\texttt{balance()}. The derived postcondition is equivalent to the manually
written postcondition of \figref{motivation:bankspec}.}\label{fig:verifying:explaints}
\end{figure}

\figref{design:twostate} shows a suitable coupling invariant for our \texttt{Bank} struct. It
states that (across each function it applies to) the change in the value of
\texttt{balance(a\_id)} should correspond to the change in resource
\texttt{Money(a\_id)} for all account identifiers \texttt{a\_id}.
To demonstrate the implications of this invariant, we can consider how it applies to
the \texttt{deposit} function (depicted graphically in \figref{verifying:explaints}).

The function's specifications using \texttt{resource} assertions prescribe the
resources to be transferred into and out of the function call; in this example,
none for the (empty) precondition and \transfers{Money(acct\_id), amt} for the
postcondition. This is elaborated (automatically) to a complete pointwise
specification of the effects on all \rtype{}s, making explicit the cases of
\emph{no} transfer via \eg{} \texttt{holds(Money(a\_id) == 0)} constraints. By
conjoining the coupling invariant, this logically implies a precise
specification relating the values of \texttt{balance} calls between the
pre-state and post-state; exactly the complex postcondition that had to be
written manually for \texttt{deposit} in \figref{motivation:bankspec}. The
combination of our novel resource specifications describing only the effects
that \emph{do} happen, our native resource semantics and an appropriate coupling
invariant, results implicitly in a logically-equivalent specification expressed
much more simply.

Coupling invariants apply to \emph{all} functions that take a (mutable)
reference to a \texttt{Bank} (in particular, they cannot be forgotten/ignored
for a function; they define a constraint on operations on this type in general).
\figref{design:transferrspec} shows the specifications for the
\texttt{transfer} function implemented using our methodology; unlike the
original version in \figref{motivation:transferspec}, the specification does not
require explicit frame conditions, and (with the exception of the extra
precondition \texttt{from != to}) is a simple composition of the corresponding
specifications of \texttt{withdraw} and \texttt{deposit} from
\figref{design:rspec}. Important resource notions such as non-duplicability and
a \zgout{built-in} notion of resource amount are built-in: in short, our new methodology
addresses the three main specification issues detailed in \secref{motivation}.

\begin{figure}[h]
\begin{rust}
#[requires(from != to)]
#[requires(~\transfers{Money(from), amt}~)]
#[ensures(~\transfers{Money(to), amt}~)]
fn transfer<B: Bank>(bank: &mut B, from: AcctId, to: AcctId, amt: u128) { ... }
\end{rust}
\caption{The specification of \texttt{transfer()}, described in terms of the resource
\texttt{Money}.}\label{fig:design:transferrspec}
\end{figure}

\zg{We note that the coupling invariants in our methodology are more lightweight
than classical type invariants, or invariants in substructural logics such as
Iris\scite{iris}; they are not required to be transitive or reflexive.
Furthermore, they are not required to hold (and cannot be assumed) within a
function body. Rather, our invariants are interpreted simply as postconditions
on affected functions and can be elaborated via a straightforward syntactic
translation.}

\zg{The coupling invariants in our methodology depend on the function
specifications summarising the overall effect of their function's resource
operations. Therefore, to extend our methodology to support a concurrency model
where resources can be transferred between threads, such transfers would need to be
encoded in the function specification. Otherwise, if a resource appeared to be
destroyed (when in fact it was transferred), the coupling invariant would
require a change to program state that could not be justified. We expect that
extending our methodology to support such operations would be straightforward.}

%% file: implementation.tex
\section{Implementation}\label{sec:implementation}

\newcommand{\den}[1]{\left\llbracket{}#1\right\rrbracket{}}
\newcommand{\denc}[3]{\den{#1}_{#2}^{#3}}
\renewcommand{\to}{\rightsquigarrow}
\newcommand{\prelabel}{\texttt{pre}}
\newcommand{\oldctx}{\ensuremath{o}}
\newcommand{\oldctxold}[1]{\textit{old}(#1)}
\newcommand{\oldctxcur}[1]{\textit{cur}(#1)}
\newcommand{\oldctxundef}{\oldctxcur{l_{\epsilon}}}
\newcommand{\denctx}{\ensuremath{c}}
\newcommand{\inden}{\texttt{+}}
\newcommand{\outden}{\texttt{-}}
\newcommand{\stdden}{\epsilon}
\newcommand{\pred}{\ensuremath{p}}
\newcommand{\predcons}{\ensuremath{P}}
\newcommand{\rescons}{\ensuremath{R}}
\newcommand{\pureexp}{\ensuremath{e}}
\newcommand{\trm}{\ensuremath{t}} % Prusti Terms
\newcommand{\impureexp}{\ensuremath{I}}
\newcommand{\typ}{\ensuremath{T}}
\newcommand{\stmt}{\ensuremath{s}}
\newcommand{\var}{\ensuremath{x}}
\newcommand{\permof}[1]{\texttt{perm(}#1\texttt{)}}
\newcommand{\holdsof}[1]{\mathtt{holds}(#1)}
\NewDocumentCommand{\oldof}{O{\ensuremath{l}} O{\pureexp}}{\texttt{old[}#1\texttt{](}#2\texttt{)}}
\newcommand{\transfersof}[1]{\mathtt{resource}(#1)}

We implemented our verification technique as an extension of the Rust program
verifier \prusti. \prusti{} verifies Rust code by encoding the program and
specifications into the intermediate verification language
Viper~\scite{muller2016viper}, which supports permission-based reasoning using a
variant of implicit dynamic frames~\scite{smans2009implicit}. Although \prusti{}
does not allow users to directly access Viper's permission primitives in
specifications, it uses Viper permissions to encode the guarantees provided by
Rust's ownership model\zg{; this enables heap framing without requiring user
annotations}. Our methodology uses Viper's permission-based reasoning
capabilities to extend Prusti with support for our novel resource reasoning
technique.

\zg{We note that our methodology is not specific to Rust or Viper. In
particular, although the Rust borrow-checker enables automatic coarse-grained
framing of the heap, our methodology does not assume these properties. Rather,
this implementation demonstrates our methodology's ability to layer user-defined
resources on top of this existing coarse-grained framing.

Although the usage of Viper's resource primitives simplifies our implementation,
our methodology does not require resources to be available in the underlying
program logic. Viper tracks resources implicitly and transfers them via proof
steps; however, this could also be done in \eg{} Dafny by tracking resources
explicitly with a mutable map \footnote{In fact, the Viper verifier Carbon
performs such a translation to Boogie.}. These two representations of resources
have been shown to be equivalent for traditional separation
logics\scite{ParkinsonSummers12}.}

\subsection{The Viper Intermediate Verification Language}\label{sec:implementation:viper}

Viper is an imperative intermediate verification language, in the spirit of
Boogie~\scite{boogie} and Why3~\scite{why3}. Program verifiers are built to
translate source language (in our case, Rust) verification problems into Viper
programs, which can then be checked by a Viper verifier.

A Viper program consists of a set of methods; Viper methods (whose bodies are
statements) are similar to functions in an imperative language. Viper statements
include variable assignments and method calls, as well as \texttt{assume} and
\texttt{assert} statements that are interpreted in the standard way. Viper also
includes \texttt{label} statements: the statement \texttt{label $l$} associates
the point in the method where the statement appears with the label $l$. The
labels are used in Viper's \texttt{old} expressions: the expression
\oldof[$l$][$e$] refers to the value of the expression $e$ at the program point
$l$.

Viper supports separation-logic style reasoning: its language and logic has its own notion of resource state, used to track a variety of \emph{resource assertions} \cite{muller2016viper}. For the purposes of this paper, the
only relevant Viper resources are \emph{\predinstfull{}s}. Viper allows the declaration of custom
\emph{abstract predicates} using the syntax
\texttt{predicate}~$\predcons(\var_{1}: \typ_{1}, \ldots, \var_{n}: \typ_{n})$,
where $\predcons$ is the predicate name, and
$\var_{1}: \typ_1, \ldots, \var_{n}: \typ_{n}$ are the names and types of its arguments. Each instantiation
$\predcons(\pureexp_{1}, \ldots, \pureexp_{n})$ of the
predicate is treated as a type of resource to be tracked.

Expressions $\mathtt{acc}(p, \mathit{amt})$ denote an amount
$\mathit{amt}$ of the \predinst{} $p$. %, where $\mathit{amt}$ is a Viper
%expression of the type \texttt{Perm} (the type of rational numbers).
%
Viper statements \texttt{inhale}~$\pureexp$ and \texttt{exhale}~$\pureexp$
add/remove resources described by \texttt{acc} expressions in $\pureexp$; for non-resource expressions they behave as assume/assert statements. Removing (via \texttt{exhale}) resources not in the state causes a verification error.
%
%if the Viper state does not contain
%sufficient amount of resource to remove the resources described by $e$, then the
%verifier raises an error.
%
Viper also supports \texttt{perm()} expressions that query an amount of
resource: an expression $\permof{p}$ refers to the amount of the
\predinst{} $p$.

\subsection{Encoding of Resources into Viper}\label{sec:implementation:encoding}

An overview of our translation from our resource specifications into Viper is shown in \figref{implementation:encoding}.
The (overloaded) syntax $\den{}$ denotes the translation of Prusti declarations, Rust
statements, and Rust types into Viper. We omit the translation of types, which is unchanged from the prior Prusti encoding \cite{prusti}. Prusti's encoding of pre-existing features is unchanged, except for adding additional \texttt{label}
statements (used for encoding our $\holds{}$ expressions).

The syntax $\denc{}{\denctx}{\oldctx}$ denotes the encoding of a Prusti
expression in the context \denctx, \oldctx. The context changes the way that
\texttt{holds()} expressions are translated into Viper. The first element, $c$,
represents the \emph{current context}: either \emph{no label} $\stdden{}$, or a \emph{signed label} $\outden{}l$
or $\inden{}l$ (denoting whether we are currently removing or adding resources). The second element, $o$, represents the \emph{old context}
taking the form of either $\oldctxold{l}$ or $\oldctxcur{l}$. These contexts prescribe
where (with respect to which existing label) generated \texttt{perm()} expressions should be evaluated and how they should be composed. For example, the
context $+l', \oldctxold{l}$, is used to encode \texttt{old(holds())}
expressions that occur in the postcondition of a called function, and are
computed by taking the difference of the \texttt{perm()} expressions taken
before and after exhaling the function precondition, corresponding to labels $l$ and $l'$ respectively.

\begin{figure}[t]
\begin{small}
\begin{minipage}[t]{\linewidth}
  \textbf{Declarations}\\
  \quad$\begin{array}{l}

  \den{\mathtt{\#[resource\_kind]~struct~}\rescons(\typ_{1}, \ldots, \typ_{n});}
  \to
  \mathtt{predicate~}\rescons(\texttt{arg1}:\den{\typ_{1}}, \ldots, \texttt{argn}:\den{\typ_{n}})\\

     \\

  \den{
    \begin{array}{l}
      \mathtt{\#[requires}(f_\mathit{pre})\mathtt{]}\\
      \mathtt{\#[ensures}(f_\mathit{post})\mathtt{]}\\
      \mathtt{fn~}f(x_1: T_1, \ldots, x_n: T_n)~\{ s_{1}; \ldots; s_{m} \}
    \end{array}
    } \to
   \begin{array}{l}
     \mathtt{method~}f(x_1: \den{T_1}, \ldots, x_n: \den{T_n})\{\\
     ~~\mathtt{inhale~}\denc{f_{pre}}{\stdden}{\oldctxundef};
     ~~\mathtt{label}~\prelabel;\\
     ~~\den{s_{1}};\ldots;\den{s_{m}}\\
     ~~\mathtt{label}~\texttt{post};~
     ~~\mathtt{exhale}~\denc{f_{post}}{\outden{}\texttt{post}}{\oldctxcur{\prelabel}}\\
     \mathtt{\}}
   \end{array}\\

  \end{array}$\\

  \textbf{Statements}\\
  \quad$\begin{array}{l}

  % produce!
  \den{\mathtt{\inhalekeyword}(\trm);} \to
  \mathtt{inhale~}\denc{\trm}{\stdden}{\oldctxcur{\prelabel}}

  \quad\quad

  % consume!
  \den{\mathtt{\exhalekeyword}(\trm);} \to
  \mathtt{exhale~}\denc{\trm}{\stdden}{\oldctxcur{\prelabel}}
  \bigskip \\

   % Function call
   \den{f(t_{1}, \ldots, t_{n});} \to
   \begin{array}{l}
     \mathtt{label}~l_{\textit{pre}};~
     \mathtt{exhale~} \denc{f_{\textit{pre}}[\var_1 := t_1, \ldots, \var_n := t_n]}{\outden{}l_{\textit{pre}}}{\oldctxundef}\\
     \mathtt{label}~l_{\textit{post}};~
     \mathtt{inhale~}
        \denc{f_{\textit{post}}[\var_1 := t_1, \ldots, \var_n := t_n]}{\inden{}l_{\textit{post}}}{\oldctxcur{l_\textit{pre}}}
   \end{array}\\

     \\
  \end{array}$

  \textbf{Expressions}\\
  \quad$\begin{array}{l}

  % Old Expressions
  \denc{\mathtt{old}(t)}{\denctx}{\oldctxcur{l}}
  \to
    \denc{t}{\denctx}{\oldctxold{l}}

  \smallskip
  \\

  % Abstract Predicate Instance
  \denc{\rescons(\trm_{1}, \ldots, \trm_{n})}{c}{\oldctx}
  \to
  \rescons(\denc{\trm_{1}}{c}{\oldctx}, \ldots, \denc{\trm_{n}}{c}{\oldctx})

  \quad\quad

  % Resource
    \denc{\transfersof{r, t}}{c}{\oldctx}
    \to
    \mathtt{acc}(\denc{r}{c}{\oldctx}, \denc{t}{c}{\oldctx})
          \bigskip
    \\

  % Holds in body (cur)
  \denc{\mathtt{holds}(r)}{\stdden}{\oldctxcur{l}}
  \to
     \mathtt{perm}(\denc{r}{\stdden}{\oldctxcur{l}})

  \quad\quad\quad\hspace{0.5em}

  % Holds in body (old)
  \denc{\mathtt{holds}(r)}{\stdden}{\oldctxold{l}}
  \to
     \oldof[l][\mathtt{perm}(\denc{r}{\stdden}{\oldctxold{l}})]
  \bigskip
     \\

  % Holds (inhale) (cur)
   \denc{\holdsof{r}}{\inden{}l'}{\oldctxcur{l}} \to
          \permof{\denc{r}{\stdden}{\oldctxcur{l}}}~\texttt{-}
          ~\oldof[l'][\permof{\denc{r}{\stdden{}}{\oldctxcur{l}}}]\\

  % Holds (inhale) (old)
   \denc{\holdsof{r}}{\inden{}l'}{\oldctxold{l}} \to
          ~\oldof[l][\permof{\denc{r}{\stdden{}}{\oldctxold{l}}}]~\texttt{-}
          ~\oldof[l'][\permof{\denc{r}{\stdden{}}{\oldctxold{l}}}]
  \bigskip
     \\

  % Holds (exhale) (cur)
   \denc{\holdsof{r}}{\outden{}l'}{\oldctxcur{l}} \to
          \oldof[l'][\permof{\denc{r}{\stdden{}}{\oldctxcur{l}}}]~\texttt{-}~\permof{\denc{r}{\stdden}{\oldctxcur{l}}}\\

  % Holds (exhale) (old)
   \denc{\holdsof{r}}{\outden{}l'}{\oldctxold{l}} \to
          \oldof[l][\permof{\denc{r}{\stdden{}}{\oldctxold{l}}}]\\

        \end{array}$
\caption[Encoding of Prusti resource constructs into Viper]{Encoding of Prusti resource constructs into Viper. Labels $l_{\textit{pre}}$ and $l_{\textit{post}}$ are assumed to be fresh. The label $l_\epsilon$ denotes an arbitrary fresh label that is never referenced, i.e., a placeholder label used where \texttt{old} expressions are not permitted.}\label{fig:implementation:encoding} \end{minipage}
\end{small}
\end{figure}

\subsubsection{Resources and Resource Operations}
As described previously, resource types (via their \Rtypecon{}s) are declared in Prusti as a Rust struct, annotated with the tag
\resourceann. 
We encode \rtypecon{}s as abstract predicates in Viper; the fields
of the struct are encoded as the arguments to the abstract predicate. We chose
this encoding because abstract predicates in Viper are used to model resources
in Viper's resource model.
Accordingly, we translate \transfers{} expressions directly into \texttt{acc()}
expressions in Viper. Our \inhalekeyword{} and \exhalekeyword{} statements are
encoded as \texttt{inhale} and \texttt{exhale} statements in Viper; these provide our desired semantics directly.

\holds{} expressions in the body of a Rust function are encoded as
\texttt{perm()} expressions in Viper. However, as described in
\secref{design:holdfunc}, our \texttt{holds()} expressions have a different
semantics in pre-/post-conditions (while Viper's \texttt{perm()} expressions do
not). Encoding our semantics requires a more-complex translation, using the
contexts on our translation function to calculate the correct differences
between \texttt{perm()} expressions at different points in the program.

\subsection{Coupling Invariants and Reborrowing}\label{sec:implementation:coupling}

We directly encode coupling invariants (declared with \texttt{\#[invariant\_twostate($t$)]} annotations on Rust structs)
as postconditions on the corresponding translated Viper methods. One restriction
of our current methodology is that we do not support Rust functions that
\emph{reborrow} mutable references, returning to the caller a live mutable
reference to \eg{} the internals of the \texttt{Bank}. An example of such a
function is shown in \figref{implementation:reborrow}, where a mutable
references to one of the balances is handed out to the caller. Technically, this
requires that the \emph{client code} would become responsible for maintaining
the \texttt{Bank}'s two-state invariant. For untrusted client code (such as
smart contracts running on a blockchain infrastructure), this should not be
relied upon, and rejecting such functions (as we currently do) is the right
approach. A similar issue and its solution has been discussed in the context of
\emph{single-state} invariants for Prusti \cite{NFM}. However, for
\emph{trusted} client code (\eg{} other verified layers of the same software
stack), we believe an adaptation of this idea to support reborrowing with our
two-state invariants, as future work (it remains to consider exactly which pairs
of states we would use to enforce our two-state invariants in a potential
extension). This feature has not been needed in practice when applying our
methodology to examples (likely because this additional reliance on client code
for correctness is often not desirable for such programs).

With this encoding in place (and implemented), we can verify Rust code specified in our methodology directly and automatically, as we evaluate in the following sections.

\begin{figure}[t]
\begin{rust}
struct Bank { balances: HashMap<AcctId, u32> }
impl Bank {
  fn get_balance_ref(&mut self, acct_id: AcctId) -> &mut u32 {
    self.balances.get_mut(acct_id).unwrap()
  }
}
\end{rust}
\caption{A function that performs a reborrow. The variable \texttt{self} is
inaccessible at the end of the function call and the returned reference can modify its internal state; the Bank's two-state invariant
cannot yet be re-established.}\label{fig:implementation:reborrow}
\end{figure}

%% file: casestudy.tex
\section{Case Study}\label{sec:casestudy}

In this case study, we take the core of a Rust implementation \scite{ibcrs}
of a cryptocurrency token transfer application running on the Interblockchain
Communication Protocol (IBC) \scite{ibc}. The token transfer application enables
tokens to be sent from one blockchain to another, without the need for a trusted
intermediary; a bug in the application could inadvertently cause tokens to be
destroyed or duplicated.

\subsection{The Fungible Token Transfer Application}\label{sec:casestudy:app}

The IBC token transfer specification allows tokens to be sent
between blockchains with different implementations and consensus
mechanisms. Its design requires that each chain track token ownership for
accounts on the chain, but not for those on other chains. The application has access to a ledger on each chain; allowing it to mint and burn
tokens for arbitrary accounts on that chain. It ``transfers'' a token between
chains by making ledger updates on both chains.

A \naive~token transfer implementation could operate by destroying the tokens on
one chain and creating them on another. However, the \naive~approach has two
issues. First, a chain may not have the capacity to mint some kinds of tokens,
i.e., tokens with fixed supply. Second, this approach would allow an exploit
allowing unrestricted minting of a token on one compromised chain to allow
obtaining an arbitrary amount of that token on any connected chain. In
particular, if a chain $A$ had a vulnerability allowing malicious arbitrary minting, an
attacker could mint on $A$ a token that exists on chain $B$, and then transfer
it to chain $B$.

Instead, the fungible token transfer application performs a token transfer from
chain $A$ to chain $B$ by first sending the token to a special \textit{escrow
account} on chain $A$, and then minting a \emph{voucher token} on chain $B$
(shown in \figref{casestudy:escrow}). The
denomination of the voucher token is created by prefixing the denomination of
$A$'s token with the identifier of the channel connecting the two chains.

\begin{figure}[t]
  \tikzset{chain/.style={draw, minimum width=2.5cm, rectangle, minimum
      height=2.6cm, inner sep=0.05cm }}
  \tikzset{account/.style={draw, minimum width=1.4cm, rectangle, minimum height=0.7cm }}
  \tikzset{coin/.style={draw,circle,minimum size=0.1cm  }}
  \pgfmathsetmacro{\acctspc}{0.65} % Spacing between accounts
  \pgfmathsetmacro{\acctlabelspc}{0.00}
  \pgfmathsetmacro{\chainlabelspc}{0.15}
  \begin{subfigure}[t]{0.45\textwidth}
  \definecolor{darkgreen}{rgb}{0.0, 0.5, 0.0}
  \begin{tikzpicture}

    \node[account] (chainaescrow) {};
    \node[coin,dashed,at=(chainaescrow.center)] (aescrowcoin) {$t$};
    \node[below=\acctlabelspc of chainaescrow] {$\emph{escrow}_B$};

    \node[account,below=\acctspc of chainaescrow] (chainafrom) {};
     \node[coin,at=(chainafrom.center)] (afromcoin) {$t$};
    \node[below=\acctlabelspc of chainafrom] (chainafromname) {\emph{source}};

    \draw[->,dashed,shorten <=2pt, shorten >=2pt] (afromcoin.west)
    to [bend left=60,looseness=1.2] node[midway,xshift=-0.8cm]
    {\textsc{escrow}} (aescrowcoin.west);

    \node[chain, fit=(chainaescrow) (chainafrom) (chainafromname)] (chaina) {};
    \node[above=\chainlabelspc of chaina] {Chain $A$};

    \node[chain, right=0.5cm of chaina.east] (chainb) {};
    \node[above=\chainlabelspc of chainb] {Chain $B$};

    \node[account,above=0.75cm of chainb.south] (chainbto) {};
    \node[coin,darkgreen,very thick,dashed,draw,at=(chainbto.center)] (btocoin) {$t'$};
    \node[darkgreen,right=-0.1cm of btocoin.east,yshift=-0.47cm] {\textsc{mint}};
    \node[below=\acctlabelspc of chainbto] {$\emph{dest}$};
  \end{tikzpicture}
  \caption{Operations performed when sending a native token on $A$ to another
    chain $B$. The token is escrowed on $A$ and a voucher is minted on $B$.}\label{fig:casestudy:escrow}
  \end{subfigure}
  \hfill
  \begin{subfigure}[t]{0.5\textwidth}
  \begin{tikzpicture}

    \node[account] (chainaescrow) {};
    \node[coin,dashed,at=(chainaescrow.center)] (aescrowcoin) {$t$};
    \node[below=\acctlabelspc of chainaescrow] {$\emph{escrow}_B$};

    \node[account,below=\acctspc of chainaescrow] (chainafrom) {};
     \node[coin,at=(chainafrom.center)] (afromcoin) {$t$};
    \node[below=\acctlabelspc of chainafrom] (chainafromname) {\emph{dest}};

    \draw[->,dashed,shorten <=2pt, shorten >=2pt] (aescrowcoin.east)
    to [bend left=60,looseness=1.2] node[midway,xshift=0.9cm]
    {\textsc{unescrow}} (afromcoin.east);

    \node[chain, fit=(chainaescrow) (chainafrom) (chainafromname)] (chaina) {};
    \node[above=\chainlabelspc of chaina] {Chain $A$};

    \node[chain, left=0.5cm of chaina.west] (chainb) {};
    \node[above=\chainlabelspc of chainb] {Chain $B$};

    \node[account,above=0.75cm of chainb.south] (chainbto) {};
    \node[coin,red,very thick,dashed,draw,at=(chainbto.center)] (btocoin) {$t'$};
    \node[red,right=0.00cm of btocoin.east,yshift=-0.47cm] {\textsc{burn}};
    \node[below=\acctlabelspc of chainbto] {$\emph{source}$};
  \end{tikzpicture}
  \caption{Operations performed when sending a voucher token back to its
originating chain. The voucher is burned on $B$ and the native token is
unescrowed on $A$.}\label{fig:casestudy:unescrow}
  \end{subfigure}
  \caption{Operations of the Token Transfer Application.}
\end{figure}

Intuitively, the voucher token on $B$ corresponds to the escrowed token on $A$.
In particular, when the application transfers the voucher token back to $A$, it
is burned on chain $B$, and the tokens in the escrow account in chain $A$ are
sent to the recipient (as shown in \figref{casestudy:unescrow}). The advantage
of this design is that it does not require the token transfer application to
mint or burn \emph{native} (i.e., non-voucher) tokens. Because this design
ensures that native tokens are never minted, exploits on one chain cannot use
the protocol to inflate the supply of native tokens on other chains.

When a voucher token is transferred to a chain other than its originating chain,
it is treated the same as any other token. This enables tokens to be transferred
transitively across multiple chains. For example, suppose $A$, $B$ and $C$ are
blockchains, and $B$ is connected to both $A$ and $C$. Then, a
token on $A$ can be sent to $C$ by first making a transfer from $A$ to $B$, and
sending the voucher minted on $B$ to $C$. $C$ ends up with a voucher for the
token on $B$, which itself is a voucher for the token on $A$.

\subsection{Application Architecture}\label{sec:casestudy:architecture}

The token transfer application is defined with respect to the
\texttt{BankKeeper} trait shown in \figref{casestudy:bank}. The application uses
the \texttt{BankKeeper} to manage the ledger on a particular chain; transfers
are accomplished by interacting with instances of the \texttt{BankKeeper} trait
on both chains.

\begin{figure}[t]
\begin{rust}
struct PrefixedDenom {trace_path: TracePath, base_denom: BaseDenom}
struct PrefixedCoin {denom: PrefixedDenom, amount: u32}
trait BankKeeper {
  fn send_tokens(&mut self, from: AccountId, to: AccountId, coin: PrefixedCoin);
  fn burn_tokens(&mut self, from: AccountId, coin: PrefixedCoin);
  fn mint_tokens(&mut self, to: AccountId, coin: PrefixedCoin);
}
\end{rust}
\caption{The \texttt{BankKeeper} interface}\label{fig:casestudy:bank}
\end{figure}

The struct \texttt{AccountId} identifies an account; within \texttt{BankKeeper}
these refer to local accounts (not those on other chains). The struct \texttt{PrefixedCoin} consists of a token
denomination and an integer amount; this type is used in the bank interface to
specify which token should be minted, burned, or transferred. The struct
\texttt{PrefixedDenom} refers to a token denomination: \texttt{base\_denom} is
the name of the token on the originating chain, and \texttt{trace\_path} denotes
the sequence of token transfers necessary to exchange a token of this
denomination with the one on its original chain.

\subsection{Properties of the Fungible Token Transfer
Application}\label{sec:casestudy:properties}

For our case study, we focused on verifying two key properties of the token
transfer application. We decided on the properties based on suggestions from
collaborators at Informal Systems (responsible for the code). The properties are as follows:

\begin{itemize}
\item\textbf{Two-Way Peg:} After transferring a token from one chain to
another, it should be possible to send the token back to the original account.
Performing this round-trip token transfer should not result in any balance
changes on either chain.
\item\textbf{Preservation of Supply:} For any given native token, the sum of the
native token (excluding tokens in escrow accounts) and all derived voucher tokens should remain
constant.
\end{itemize}

These properties originate from the IBC specification \scite{tokentransferspec},
although the two-way peg property we consider is stronger than the version in
the original specification. To verify the implementation, it was necessary to
make some changes to the code in order to workaround limitations of Prusti. For
example, the \texttt{TracePath} data structure is implemented using data
structures that are not fully supported by Prusti; therefore, we simplified the
definitions of such types. However, these changes are orthogonal to our
verification methodology.

\zgout{In addition, we have not yet implemented support for coupling invariants on
\emph{traits}, therefore, to verify the implementation we changed the interface to
define \texttt{BankKeeper} as a struct instead. As the expected properties and
behaviour of the \texttt{BankKeeper} remain the same, this change is not
significant, and the results of our verification are still applicable to the
original design.}

\zgedit{The token transfer application relies on implementations of the IBC core
standards~\cite{ibccoreprotocol} to coordinate communication across chains. For
example, executing code on the remote chain involves generating cryptographic
proofs, serialising message data, and routing messages appropriately. These
aspects are unrelated to the application logic itself, furthermore, verification
of these aspects would likely not benefit from our resource reasoning
methodology. Therefore, we assume that these components function correctly.}{
The token transfer application relies on an implementation of the IBC core
standards~\cite{ibccoreprotocol} to coordinate communication across chains.
We assume the correctness of this implementation, as verification
of these aspects would likely not benefit from our resource reasoning
methodology.}

\subsection{Encoding the Specification with Resources}\label{sec:casestudy:spec}

We present our specification by first explaining the \rtype{}s that are used to
model the resources in the program (\secref{casestudy:rkind}). We then
demonstrate the specifications for the \texttt{BankKeeper}
(\secref{casestudy:verifybankkeeper}) and constituent functions of the
token-transfer application (\secref{casestudy:verifyapp}). In
\secref{casestudy:verifyproperties} we present the specifications used to verify
the properties.

\subsubsection{Representing Tokens with \RType{}s}\label{sec:casestudy:rkind}

We begin our presentation by identifying appropriate \rtype{}s for use in
specifications. First, we note that the two properties above refer to different
aspects of tokens: the two-way peg property is concerned with the tokens in each
account, while the preservation of supply considers the sum of unescrowed tokens
across all accounts. The latter property does not distinguish between native and
voucher tokens; however, this distinction is relevant for the former. Therefore,
we define \emph{two} different resource constructors, each corresponding to a
different aspect of the same tokens. We declare the first \rtypeconshort{},
which is used to encode the two-way peg property, as follows:

\begin{rust}
~\resourceann{}~ struct Money(BankID, AccountId, PrefixedDenom)
\end{rust}

Resources constructed using the \rtypeconshort{} \texttt{Money} have distinct
\rtypeshort{}s if they belong to different banks, differ in denomination, or have
different owners. We\zgout{can} ensure the two-way peg property by showing that
round-trip transfers do not change the amount of \zgedit{tokens of this
\rtypeshort{}}{\texttt{Money}} in the resource state.

Proving the second property requires showing that the total amount of all
unescrowed tokens of a particular base denomination remain unchanged. For this
property, we use a different \rtypecon{}, that treats tokens as having the same
\rtypeshort{} when they are in the same bank and have the same base
denomination. We declare the \rtypeconshort{} as follows:

\begin{rust}
~\resourceann{}~ struct UnescrowedCoins(BankID, BaseDenom);
\end{rust}

As previously mentioned, each \rtypecon{} corresponds to a different view of the
same token. Therefore, operations on real tokens should affect the corresponding
resources for each view in a uniform manner. Therefore, we can define a macro to
describe these real-world operations in terms of how they affect each view of
the resource; this macro simplifies the specification, since it eliminates the
need to write resource operations for each view separately.

\begin{figure}[t]
\begin{rust}
macro_rules! transfer_money { ($bank_id:expr, $to:expr, $coin:expr) => {
  ~\transfers{Money(\$bank\_id, \$to, \$coin.denom), \$coin.amount}~ &&
  if !is_escrow_account($to) {
    ~\transfers{UnescrowedCoins(\$bank\_id, \$coin.denom.base\_denom), \$coin.amount}~
  } else { true }
}}
\end{rust}
\caption{A macro that specifies what Prusti resources are changed (i.e., either created or destroyed) in response to token operations. An operation on tokens will always change the \texttt{Money} resource, and also changes the \texttt{UnescrowedCoins} resource if the target account is not an escrow account.}\label{fig:evaluation:transfermacro}
\end{figure}

The resulting macro is presented in \figref{evaluation:transfermacro}. The macro
states that a transfer of a token should always move an instance of the
\texttt{Money} resource constructor, but should only move the corresponding
\texttt{UnescrowedCoins} resource if the account \texttt{\$to} is not an escrow
account.

\subsubsection{Specifying the BankKeeper}\label{sec:casestudy:verifybankkeeper}

We use the \texttt{transfer\_money()!} macro to annotate \texttt{BankKeeper}'s
associated functions with appropriate resource operations, as presented in
\figref{evaluation:annotatedbank}.

\begin{figure}[t]
\begin{rust}
trait BankKeeper {
  #[requires(transfer_money!(self.id(), from, coin))]
  #[ensures(transfer_money!(self.id(), to, coin))]
  fn send_tokens(&mut self, from: AccountId, to: AccountId, coin: PrefixedCoin);

  #[requires(transfer_money!(self.id(), from, coin))]
  fn burn_tokens(&mut self, from: AccountId, coin: PrefixedCoin);

  #[ensures(transfer_money!(self.id(), to, coin))]
  fn mint_tokens(&mut self, to: AccountId, coin: PrefixedCoin);
}
\end{rust}
\caption{The \texttt{BankKeeper} \zg{trait} annotated with the appropriate resource operations.}\label{fig:evaluation:annotatedbank}
\end{figure}

We then connect the resource operations to properties about the program state by
establishing coupling invariants on the \texttt{BankKeeper} function, which
describe how changes to \texttt{Money} and \texttt{UnescrowedCoins} in the
resource state correspond to changes in functions \texttt{balance} and
\texttt{unescrowed\_coin\_balance} respectively. These invariants are shown in
\figref{evaluation:invariants}.

\begin{figure}[t]
\begin{rust}
#[invariant(forall(|acct_id: AccountId, denom: PrefixedDenom|
  holds(Money(self.id(), acct_id, denom)) -
  old(holds(Money(self.id(), acct_id, denom))) ==
  self.balance(acct_id, denom) - old(self.balance(acct_id, denom))))]
#[invariant(forall(|coin: BaseDenom|
  holds(UnescrowedCoins(self.id(), coin)) -
  old(holds(UnescrowedCoins(self.id(), coin))) ==
  self.unescrowed_coin_balance(coin) - old(self.unescrowed_coin_balance(coin))))]
\end{rust}
\caption{\zgedit{Invariants}{Coupling invariants} connecting the resources \texttt{Money} and
\texttt{UnescrowedCoins} to the methods \texttt{balance()} and
\texttt{unescrowed\_coin\_balance()} respectively.}\label{fig:evaluation:invariants}
\end{figure}

\subsubsection{Verifying the Application Logic}\label{sec:casestudy:verifyapp}

We now focus our attention to the logic of the token-transfer application
itself. The application performs a token transfer from a chain $A$ to chain $B$
in two steps. The first step is performed by calling the function
\texttt{send\_fungible\_tokens} (\figref{eval:send}) on chain $A$, which
disposes of the tokens on $A$ by either escrowing or burning them. In the second
step, it effectively causes the function \texttt{on\_recv\_packet} to be called
on chain $B$, which produces the tokens by either minting or unescrowing
tokens as necessary\footnote{Technically, \texttt{on\_recv\_packet} in the
original version is a callback that is triggered when a relayer sends a message
to chain $B$, however for verification we consider this as a regular function
call. The routing of callbacks is implemented as part of the IBC core standard,
which we assume to be correct and do not attempt to verify.}.

\figref{eval:send} shows a specification and implementation of the
\texttt{send\_fungible\_tokens} function (we simplify some details for
presentation here), which performs the initial step in the cross-chain token
transfer. The arguments \texttt{port} and \texttt{channel} identify the chain
that tokens will be transferred to. The condition on line 6 checks if the token
being sent is a voucher for a token on that chain. If so, then the voucher is
burned on this chain (and the next step will unlock the corresponding token on
the other chain). Otherwise, the token is moved to an escrow account on this
chain.
\begin{figure}[t]
\begin{rust}
#[requires(transfer_money!(bank.id(), sender, coin))]
#[ensures(!coin.denom.trace_path.starts_with(port, channel) ==>
    transfer_money!(bank.id(), escrow_address(channel), coin))]
fn send_fungible_tokens<B: Bank>(bank: &mut B, coin: &PrefixedCoin,
  sender: AccountId, port: Port, channel: ChannelEnd) {
    if coin.denom.trace_path.starts_with(port, channel) { bank.burn(sender, coin);}
    else { bank.send(sender, escrow_address(channel), coin); }
}
\end{rust}
\caption{The first step of the fungible token transfer.}\label{fig:eval:send}
\end{figure}

The precondition of \texttt{send\_fungible\_tokens} specifies that calling the
function transfers the token out of the sender's account. The postcondition
\zgedit{indicates}{states} that when the token did not originate on the opposite
chain (i.e., the chain identified by \texttt{port} and \texttt{channel}), it
will be transferred to the escrow account. Due to space constraints, we do not
present the specifications of \zgedit{the second step of the
transfer}{\texttt{on\_recv\_packet()}} here; the relevant specification is
available in our supplementary material~\cite{supplementary}.

\subsubsection{Verifying the Desired Properties}\label{sec:casestudy:verifyproperties}

With the above specifications, we can then prove that the token transfer
application satisfies the properties described in \secref{casestudy:properties}.
To prove the two-way peg property, we construct a method that performs a round
trip token transfer, and show that the balances of all accounts are unchanged
after the transfer. The relevant specifications are presented in
\figref{eval:roundtrip} (the body of the method is omitted for brevity). We
verify the preservation of supply property by showing that the supply is
preserved after an arbitrary transfer; the specifications are presented in
\figref{eval:preservespec}. The specifications of both properties are expressed
in terms of the functions \texttt{balance} and
\texttt{unescrowed\_coin\_balance} respectively.

\begin{figure}[t]
\begin{rust}
#[ensures(forall(|acct_id2: AccountId, denom: PrefixedDenom|
    bank1.balance(acct_id2, denom) == old(bank1).balance(acct_id2, denom)))]
#[ensures(forall(|acct_id2: AccountId, denom: PrefixedDenom|
    bank2.balance(acct_id2, denom) == old(bank2).balance(acct_id2, denom)))]
fn round_trip<B: Bank>(bank1: &mut B, bank2: &mut B, coin: &PrefixedCoin,
    sender: AccountId, receiver: AccountId, ...) { ... }
\end{rust}
\caption[The relevant specifications for the two-way peg
property]{The relevant specifications for the two-way peg
property. The function \texttt{round\_trip()} performs a token transfer from
account \texttt{sender} to \texttt{receiver}, and then performs the same
transfer in the opposite direction.}\label{fig:eval:roundtrip}
\end{figure}

\begin{figure}[t]
\begin{rust}
~\label{line:eval:preservespecstart}~#[ensures(forall(|c: BaseDenom|
      bank1.unescrowed_coin_balance(c) + bank2.unescrowed_coin_balance(c) ==
  old(bank1.unescrowed_coin_balance(c) + bank2.unescrowed_coin_balance(c))))]~\label{line:eval:preservespecend}~
fn transfer<B: Bank>(bank1: &mut B, bank2: &mut B, coin: &PrefixedCoin,
    sender: AccountId, receiver: AccountId, ...) { ... }
\end{rust}
\caption[The relevant specifications for the supply preservation property]{The relevant specifications for the supply preservation property. The
function \texttt{transfer()} performs a token transfer from account
\texttt{sender} to \texttt{receiver}.}\label{fig:eval:preservespec}
\end{figure}

In conclusion, we've shown that our methodology can be applied to a real-world
resource-manipulating program. Using our methodology, we were able to prove two
important properties about the token-transfer application: that it maintains a
two-way peg and preserves total token supply. Our technique allows us to verify
the desired properties in a straightforward manner by describing resource
operations directly, without having to write frame conditions.

%% file: evaluation.tex
\section{Evaluation}\label{sec:evaluation}

\newcommand{\token}[1]{\texttt{TT}-#1}
\newcommand{\nft}[1]{\texttt{NFT}-#1}

To evaluate our technique, we consider how specifications written using our
methodology compare to \zgout{alternative} specifications written in terms of program
states. We consider \zgedit{four}{three} research questions:

\begin{enumerate}[label=\textbf{RQ\arabic*}]
    \item (\emph{Conciseness}): Are our specifications \emph{smaller} than the alternative?
    \item (\emph{Complexity}): Are our specifications \zg{(syntactically)} \emph{simpler} than the alternative?
    \item (\emph{Verification Time}): Do our specifications \emph{require
        more time to verify} than the alternative?
\end{enumerate}

To answer these questions, we compare the specifications we developed in
\secref{casestudy} to alternative specifications for the same implementation,
written without using our methodology.
\zgout{In addition, we also apply our methodology to a simplified implementation
of the IBC Non-Fungible Token (NFT) Transfer application, again comparing the
results to an alternative version of the specification.}
\zg{In addition, we perform a similar comparative evaluation for three other programs:}
\subsubsection*{\zg{IBC Non-Fungible Token (NFT) Transfer}}
The NFT transfer application\scite{nfttransferspec} has a similar architecture to
the token transfer application. We wrote the implementation ourselves, based on
pseudocode in the specification. \zgout{We are not aware of a functioning Rust
implementation of the protocol; ultimately we intend to develop our prototype
into a verified reference implementation.}A key difference in the application of
our methodology between the two protocols is that we use resources to model the
\emph{permission} to change ownership of a token, rather than modelling the
token itself.
\subsubsection*{\zg{\erc{721} Implementation}}
\zg{We verified a smart contract implementing NFT functionality (satisfying
the \erc{721} specification\scite{erc721}), written using the \ink{}
library\scite{ink}. This smart contract was written by the \ink{} developers, and
is presented on the project's homepage as an example application\footnote{https://use.ink/}. As in the NFT
transfer application, we use resources to model permissions. We also use
resources to represent token balances (\ie{} the number of tokens owned by an
account). In addition, the \erc{721} protocol allows an account to approve
other accounts to transfer their tokens on their behalf, and we use resources to
represent such approvals as well.}
\subsubsection*{\zg{\psp{22} Implementation}} \zg{Finally, we verified a smart
contract, also written with \ink{}, that implements fungible token functionality
on the Aleph Zero blockchain\scite{alephzero}. As of February 2024, this
contract was used by the Panorama Swap decentralised
exchange\scite{panoramaswap}. Similar to the \erc{721} protocol, \psp{22} allows
accounts to grant allowances to other accounts, allowing them to spend their
tokens, up to an allowed amount. In addition to using resources to model token
balances, we also use them to to model these allowances.}

\zg{In total, we wrote specifications for (and verified) all \numcomparisons{}
resource-related functions across the four applications. We note that it crucial
for these functions to be correct: bugs in the functions could result in a loss
(or inflation) of the digital assets they manipulate. In fact, in the process of
performing the evaluation we identified a bug in the \erc{721} implementation
that could incorrectly cause a user's balance to be decremented (this bug caused
verification to fail, as expected). We corrected the bug, and submitted a pull
request to the developers, which was accepted\footnote{<Github PR link redacted
  for double-blind submission>. We also identified a separate bug unrelated to
  resources (but necessary for verification) where a panic could occur in this
  same codebase. The corresponding PR was accepted as well: <Redacted>.}.}

\zg{The specifications and implementation for all applications included in our
evaluation are included in our supplementary material~\scite{supplementary}}. We
now consider each research question in order\zgout{, describing how we compared the
specifications and presenting our results}, and discuss \zgedit{the}{our overall} results in
\secref{evaluation:analysis}.

\subsection{Conciseness}

To measure conciseness, we compared the number of lines of code used to specify
resource-related operations in both specifications. Specification lines
unrelated to resource operations are identical in both versions and therefore
are not considered in the comparison. Our results are presented in Table
\ref{table:evaluation:complexity}. On average, specifications encoded using
resource reasoning require \comparisonlineratio{} fewer lines of code compared
to the alternative specification without resources. Of the \numcomparisons{}
functions, using resource specifications required fewer lines of code for
\numbettercomparisons{} and the same number of lines for
\numsamelinecomparisons{}.

Resource specifications required more lines of code for the remaining
\numtowords{\numworsecomparisons} functions. These functions were part of the
NFT transfer application, where resources describe \emph{permission} to modify
ownership of an NFT, rather than NFT ownership directly. Therefore, it is also
necessary to describe \emph{how} the ownership changes, \ie{} specifying which
account becomes the new owner of the token. As a result, these tokens must be
referred to twice in the specification with resources, but only once in the
specification without resources. Because the application allows transferring
multiple NFTs at once, specifying the tokens that are transferred by a function
requires more complex specifications than in the other specifications. In this
case, the overhead of specifying permissions in the resource specification
outweighs the increased specification length due to frame conditions in the
alternative version.
\zgout{Table
\ref{tab:evaluation:loc}. provides a comparison between the number of
specification lines required for both specification versions. In total, the
specifications concerned with resource properties (including invariants and
helper macro definitions) consist of lines of code. Encoding the
equivalent specifications without resources required lines.}

\begin{table}[h]
\centering
\begin{footnotesize}
\begin{tabular}{clcccccc}
\toprule
\multirow{2}{*}{\textbf{Application}} & \multirow{2}{*}{\textbf{Usage}} &
\multicolumn{3}{c}{\textbf{Without Resources}} & \multicolumn{3}{c}{\textbf{With Resources}} \\
\cmidrule(lr){3-5} \cmidrule(lr){6-8}
& & Size & Depth & \# Lines & Size (Diff.) & Depth (Diff.) & \# Lines (Diff.) \\
\midrule
\input ast_table.tex
\bottomrule
\end{tabular}
\end{footnotesize}
\caption{Comparison between the \zg{lines of code and} syntactic complexity of
specifications written with and without resources. Each column considers the
\zgedit{ASTs of the specification expressions}{specifications} for the function
in that row. \textbf{Size} refers to the total number of nodes in the ASTs.
\textbf{Depth} refers to the height of the tallest AST. \zg{\textbf{\# Lines} is the
count of lines in the specification. The percentages in the (Diff.) columns is
the ratio of the corresponding value for the specification written with
resources to the value for the specification written without
resources.}\zgout{\textbf{Uniq.} refers to the number of distinct node types
occurring in the ASTs.}}\label{table:evaluation:complexity}
\end{table}

\zg{To use resources in our specifications, we must define the resources
themselves, as well as associated coupling invariants. Table
\ref{table:evaluation:resourceannotations} shows the lines of code required for
resource annotations in the examples we consider consider. We note that they are
essentially a one-time cost, as they only need to be written once, regardless of
the number of functions in the program.}

\begin{table}[t]
\centering
\begin{footnotesize}
\begin{tabular}{lcc}
\toprule
\textbf{Application} & \textbf{Resource Definitions} & \textbf{Coupling Invariants} \\
\midrule
\input resource_table.tex
\bottomrule
\end{tabular}
\end{footnotesize}
\caption{The number of lines of code defining resource kinds and coupling
invariants for each application.}\label{table:evaluation:resourceannotations}
\end{table}

\subsection{Syntactic Complexity}

More concise specifications are not necessarily simpler or more desirable. For
example, a longer specification may be preferable to a shorter one if the latter
involves complex nesting of conditionals, implications, and quantifiers. To
evaluate syntactic complexity, we considered the AST nodes of \zg{the (parsed)
specification expressions.} We quantify the complexity of an AST based on
\zg{the total number of nodes it contains and its maximum depth. As shown in
Table~\ref{table:evaluation:complexity}, the specifications written using our
methodology are syntactically simpler in most cases: of the \numcomparisons{}
functions, \numbetterdepth{} of them have specifications with smaller AST depth
in our methodology, \numbettersize{} have smaller size. On average, the
specification ASTs have a \comparisonsizeratio{} smaller size and
\comparisondepthratio{} smaller depth.}

\subsection{Verification Time}

\begin{table}[t]
\centering
\begin{footnotesize}
\begin{tabular}{lcc}
\toprule
\textbf{Application}
  & \textbf{Without Resources (mean / sd)}
  & \textbf{With Resources (mean / sd)} \\
\midrule
\textbf{Token Transfer} & 61.83s / 0.27s & 78.37s / 0.18s \\
\textbf{NFT Transfer} & 123.05s / 1.70s & 140.34s / 0.09s \\
\textbf{\erc{721}} & 62.72s / 0.10s & 98.16s / 0.19s \\
\textbf{\psp{22}} & 36.24s / 0.26s & 51.51s / 0.09s \\
\bottomrule
\end{tabular}
\end{footnotesize}
\caption{Comparison of verification time for the specifications written in the
different styles. Results are presented for five runs of the
verifier.}\label{tab:evaluation:time}
\setlength{\belowcaptionskip}{0pt}
\end{table}

We compared the runtime of specifications written using our methodology\zgedit{,
and}{ to} the alternate version. For each version, we performed five runs, all
runs were performed a 10-core Apple M1 Max. Our results are presented in
Table~\ref{tab:evaluation:time}\zgedit{. The verification time for the NFT
transfer application is similar for both versions (59.88s vs 57.04s). There is a
larger difference \wrt{} token transfer application: the specification using
resources is 27\% slower (83s vs 65.12s)}{; overall, the specifications using
resources were 35\% slower than those without}.\zgedit{This is most likely due
to the overhead of casting between the types used to represent resource amounts
and the types representing bank balances in the underlying Viper encoding:
changing the specification to instead express balance using Viper's permission
amount types (i.e., rational numbers as opposed to integers) eliminates the
difference (resulting in timings of 88.74s vs 90.38s respectively)}{This is most
likely due to two reasons. Currently, our encoding into Viper requires
type-casting between the types used to represent resource amounts (i.e.,
rational numbers) and other numeric values in the code (e.g. token balances,
which are represented using integers).} As future work, we \zgedit{believe it
could be possible to}{could try to} reduce the performance overhead associated
with such casts.

\zg{Second, in some cases, the postconditions derived from coupling invariants
are less efficient to verify compared to specifications that would be written
manually. In particular, for methods that do not produce or consume any
resources, coupling invariants will introduce postconditions in the form of
pointwise assertions on the program state, asserting that everything is
unchanged. In contrast, the hand-written postcondition could potentially be
simpler. For example, in the case where a map \texttt{balanceMap} type tracks
account balanceMap in an application, the postcondition \texttt{balanceMap ==
old(balanceMap)} introduces less verification overhead than the automatically
generated invariant postcondition \texttt{forall(|a: AcctId| balanceMap.get(a)
== old(balanceMap.get(a)))}.}

\subsection{Analysis / Conclusion}\label{sec:evaluation:analysis}

Our evaluation shows that specifications written in our methodology compare
favourably to specifications written in terms of program states. Our
specifications require fewer lines of code and \zg{are} syntactically simpler.
\zgout{Specifications written using our methodology are also easier to
interpret: this is demonstrated by the observation that specifications written
in terms of program states must express many non-essential properties in their
specifications (\eg{} frame conditions).} Although our specifications
\zgout{sometimes}require more time to verify, the increase in time is a
reasonable trade-off to make for the simpler specifications.\zgout{Furthermore,
our evaluation indicates that the increase is due to the behaviour of the
underlying Viper verifier rather than a fundamental consequence of using our
methodology.}

\zg{We hypothesise that, in additional to being more concise and syntactically
simpler, specifications written using our methodology are easier to understand,
because they leverage our common-sense notion of resources. Evaluating this
claim, \eg{} via a user study, is a potential area of future work.}

%% file: ast_table.tex
\multirow{7}{*}{\textbf{Token Transfer}} & \texttt{burn()} & 45 & 8 & 27 & 5 (-89\%) & 3 (-63\%) & 1 (-96\%) \\
& \texttt{mint()} & 37 & 8 & 26 & 5 (-86\%) & 3 (-63\%) & 1 (-96\%) \\
& \texttt{send()} & 136 & 11 & 36 & 10 (-93\%) & 3 (-73\%) & 2 (-94\%) \\
& \texttt{send\_fungible\_tokens()} & 34 & 6 & 13 & 20 (-41\%) & 6 (-0\%) & 3 (-77\%) \\
& \texttt{on\_recv\_packet()} & 45 & 6 & 19 & 20 (-56\%) & 5 (-17\%) & 6 (-68\%) \\
& \texttt{send\_preserves()} & 47 & 7 & 16 & 57 (+21\%) & 6 (-14\%) & 16 (-0\%) \\
& \texttt{round\_trip()} & 53 & 6 & 19 & 44 (-17\%) & 5 (-17\%) & 15 (-21\%) \\
\midrule
\multirow{7}{*}{\textbf{NFT Transfer}} & \texttt{burn()} & 31 & 9 & 8 & 14 (-55\%) & 4 (-56\%) & 2 (-75\%) \\
& \texttt{mint()} & 39 & 9 & 9 & 22 (-44\%) & 4 (-56\%) & 3 (-67\%) \\
& \texttt{transfer()} & 33 & 9 & 8 & 22 (-33\%) & 4 (-56\%) & 3 (-63\%) \\
& \texttt{create\_or\_update\_class()} & 16 & 7 & 4 & 0 (-100\%) & 0 (-100\%) & 0 (-100\%) \\
& \texttt{send\_nft()} & 108 & 11 & 24 & 115 (+6\%) & 9 (-18\%) & 26 (+8\%) \\
& \texttt{on\_recv\_packet()} & 185 & 13 & 45 & 206 (+11\%) & 11 (-15\%) & 52 (+16\%) \\
& \texttt{round\_trip()} & 126 & 10 & 35 & 144 (+14\%) & 10 (-0\%) & 37 (+6\%) \\
\midrule
\multirow{12}{*}{\textbf{ERC-721}} & \texttt{set\_approval\_for\_all()} & 72 & 9 & 10 & 31 (-57\%) & 7 (-22\%) & 3 (-70\%) \\
& \texttt{approve()} & 66 & 9 & 10 & 28 (-58\%) & 5 (-44\%) & 5 (-50\%) \\
& \texttt{transfer()} & 110 & 12 & 17 & 98 (-11\%) & 9 (-25\%) & 14 (-18\%) \\
& \texttt{transfer\_from()} & 104 & 11 & 17 & 98 (-6\%) & 7 (-36\%) & 15 (-12\%) \\
& \texttt{mint()} & 88 & 10 & 13 & 46 (-48\%) & 7 (-30\%) & 5 (-62\%) \\
& \texttt{burn()} & 83 & 10 & 13 & 39 (-53\%) & 8 (-20\%) & 13 (-0\%) \\
& \texttt{transfer\_token\_from()} & 104 & 11 & 17 & 114 (+10\%) & 8 (-27\%) & 15 (-12\%) \\
& \texttt{remove\_token\_from()} & 76 & 9 & 14 & 28 (-63\%) & 4 (-56\%) & 6 (-57\%) \\
& \texttt{add\_token\_to()} & 84 & 10 & 13 & 54 (-36\%) & 5 (-50\%) & 7 (-46\%) \\
& \texttt{approve\_for\_all()} & 72 & 9 & 10 & 31 (-57\%) & 7 (-22\%) & 3 (-70\%) \\
& \texttt{approve\_for()} & 66 & 9 & 10 & 34 (-48\%) & 5 (-44\%) & 6 (-40\%) \\
& \texttt{clear\_approval()} & 64 & 8 & 8 & 19 (-70\%) & 5 (-38\%) & 4 (-50\%) \\
\midrule
\multirow{5}{*}{\textbf{\psp{{22}}}} & \texttt{transfer()} & 86 & 11 & 20 & 42 (-51\%) & 6 (-45\%) & 4 (-80\%) \\
& \texttt{transfer\_from()} & 161 & 11 & 40 & 111 (-31\%) & 8 (-27\%) & 19 (-53\%) \\
& \texttt{approve()} & 49 & 9 & 14 & 43 (-12\%) & 5 (-44\%) & 14 (-0\%) \\
& \texttt{increase\_allowance()} & 51 & 9 & 10 & 20 (-61\%) & 5 (-44\%) & 4 (-60\%) \\
& \texttt{decrease\_allowance()} & 60 & 12 & 12 & 20 (-67\%) & 6 (-50\%) & 3 (-75\%) \\

%% file: resource_table.tex
\textbf{Token Transfer} & 9 & 25 \\
\textbf{NFT Transfer} & 6 & 10 \\
\textbf{ERC-721} & 4 & 37 \\
\textbf{PSP-22} & 2 & 14 \\

%% file: related.tex
\section{Related Work}\label{sec:related}

Effect systems\scite{lucassen88effects} extend a type system to include the
side-effects via effect types, which typically over-approximate the side-effects an expression is
allowed to perform. Our resource specifications can be seen as similar to a kind of effect, but are \emph{precise}, and apply only to changes to our ghost resource state. This distinction (and indirection via our coupling invariants) is crucial to obtaining strong frame properties while still writing local specifications.

\zg{Linear type systems\scite{lineartypes} enable values to be treated as
resources: values of a linear must be used exactly once. Bounded linear types
allows resource amounts to be modelled by a semiring\scite{ghica2014bounded}.
Tracking resource usage with linear types requires associating resource amounts
with values in the program, tracking resources in this way may require changes
to the program (consuming a resource must coincide with the disposal of a
linearly-typed value, for example). In contrast, our methodology is more
flexible, as resources are tracked via ghost state.}

Separation logic\scite{reynolds2002separation} enables verification of
heap-manipulating programs using local reasoning: assertions describe only the
relevant part of the heap, rather than the heap as a whole. Our resource
reasoning technique takes clear inspiration from this idea of local reasoning,
although our technique intentionally avoids explicit resource reasoning about
the concrete program state. Extensions of separation logic facilitate
abstraction with user-defined predicates \scite{abstractpredicates}, but we note
that an abstraction of our \eg{} \texttt{Bank} interface via abstract predicates
would typically suffer from the same need for frame conditions about \emph{how}
the internals change over a side-effectful function. Our \texttt{holds}
construct, which allows introspection on the local resource state, does not have
an analogue in separation logic. However, the Viper intermediate
language\scite{muller2016viper} supports resource reasoning and resource
introspection via the \texttt{perm()} expressions explained in
\secref{implementation}. In contrast to our \texttt{holds} construct, Viper's
\texttt{perm()} expressions cannot be used directly in method pre- and
postconditions in a generally-sound way; their semantics is not consistent
between how a caller and callee interpret their meaning \cite{muller2016viper}.
Being an intermediate language, Viper provides features powerful enough to
encode our reasoning principles, but places the burden on the user to use these
features soundly. In contrast, our \texttt{holds} expressions can be used freely
in pre- and post-conditions without fear of unsoundness, and the trickier parts
of our encoding (\secref{implementation}) take care of a correct mapping to
Viper automatically.

\zg{Iris~\cite{iris} is a concurrent separation logic that allows user to define
resources using ghost state, and establish invariants that tie ghost and program
state. While Iris is more expressive than our approach (and potentially useful
as a tool to prove its soundness), it is not clear how it could be integrated
into an existing program verifier. In contrast, extending an existing program
verifier with our methodology is straightforward due to our lightweight
approach.}

There is substantial prior work focused on verification of smart contracts themselves~\cite{DBLP:conf/isola/AhrendtB20}\cite{twovyper}\cite{mohajerani2022modeling}: while not the specific focus of our work, these are clearly resource-manipulating
programs (used to handle cryptocurrency assets). The
verification tool 2Vyper \scite{twovyper} is closest to our work: it provides its own resource reasoning via effect clauses for smart
contracts in the Vyper language: these can specify possible resource transfers, and users can also
define invariants that connect the resource state to the runtime state. However, there are several technical differences with our work: unlike our system, 2Vyper's resource specifications describe approximations (in the style of effect systems) of a function's behaviour, and over representations of the entire resource state; there is no way to partition the resource state into the local part that a function is concerned with, which is what eliminates heavyweight frame conditions in our work.

2Vyper's effect-clauses consist of a multiset of the operations that will occur
in their execution, but \zgout{these} only refer to \zgout{the} resource operations performed
directly by the function \zgout{itself}: because 2Vyper considers interactions with
unverified external code, it is impossible to reason in general about external
resource operations. In contrast, because we do not allow untrusted external
calls, \zgout{the} specifications in our methodology effectively summarise the
resource operations that occur within a method call. We have not \zgout{yet} applied our
technique to verify interactions with untrusted code; doing so could be
interesting future work.

Ahrendt and Bubel verify Solidity contracts with a proof technique centred around two-state invariants  \cite{DBLP:conf/isola/AhrendtB20}. They show that \emph{differences} between \emph{wei} amounts (the built-in currency) are powerful for expressing such invariants, similar to the encoding of our \texttt{holds} feature into Viper. They address untrusted code and security properties while we do not; on the other hand, their work does not support custom notions of resource, build in resource-like properties, or address data structure framing; their technique is concerned specifically with currency in Solidity.

Other prior work has focused on verification of smart contracts by modelling them as
extended finite state machines \scite{mohajerani2022modeling}. Our approach is
more general, as it is not limited to the domain of smart contracts and does not
assume any particular program architecture.

The specification language Chainmail \scite{chainmail} enables user-defined
invariants using holistic specifications, which can be used to enforce a wide
range of security-related properties, including some related to resources in the
program. For example, it is possible to define an invariant that ensures any
change in the balance of a particular account is associated with a call to
\texttt{deposit} referencing that account. Specifications in Chainmail must
still ultimately be phrased in terms of program states; leading to less direct
and more-complex specifications. In contrast, our methodology, which treats
resources as first class, allows specifications concerning resource operations
directly.

Various smart contract languages provide first-class support for resources. The
Move language\scite{movelang}, originally developed by Facebook for the Diem
blockchain, supports first-class resources that are implemented with linear
types. Obsidian\scite{obsidian} uses both linear types and typestate to prevent
bugs related to improper handling of resources. Flint\scite{flint} supports asset
types that encapsulate unsafe operations and provide a safe interface.
None of these support static verification concerning resource quantities, or built-in rules to enforce \eg{} that are by-default preserved.

%% file: conclusion.tex
\section{Conclusion}\label{sec:conclusion}

In this paper, we examined the challenges encountered when writing
specifications of resource-manipulating programs in terms of program state. When
using a modular verifier, this approach requires users to explicitly write frame
conditions in specifications, and these frame conditions make specifications
lengthier and harder to interpret. Furthermore, such specifications do not
easily compose, and are not expressive enough to rule out certain kinds of
resource-related bugs.

The root cause of these issues is the semantic gap between our
\zgout{high-level, intuitive} expectations of how resources behave, and the
language used to write specifications\zgout{ in the code}. Therefore, we
developed a new methodology to support resource reasoning within specifications,
thereby narrowing the semantic gap.

The methodology we developed extends a program verifier with a first-class
notion of resources, without requiring support for resources in the source
language. Instead, we allow users to define coupling invariants to connect
resource operations to program state. These invariants are checked by the
verifier, allowing users to describe the behaviour of their program in terms of
resource operations rather than as relations on program states.

We implemented our methodology as an extension to the program verifier Prusti,
and evaluated \zgedit{our design}{it} by using our extended version of Prusti to verify
\zgout{a} real-world resource-manipulating program\zg{s}. Our evaluation shows
that, compared to a standard Prusti specification written in terms of program
states, specifications written using our methodology are more
concise\zgedit{,}{ and} syntactically simpler\zgout{, and easier to understand}.

For future work, we would like to extend our methodology to facilitate
interactions with untrusted or external code, as such interactions are typical
in dealing with smart contracts. In particular, enforcing coupling invariants
for functions that reborrow could increase the applicability of our technique.
\zgout{Finally, we could consider applying our methodology to reason about resources
within a program, such as locks, database connections, or file handles.}